\documentclass[aps,prx,twocolumn,superscriptaddress]{revtex4-1}
 \pdfoutput=1
\usepackage{amsmath}
\usepackage{amssymb}
\usepackage{graphicx}
\usepackage{enumerate}
\usepackage[colorlinks=true]{hyperref}
\def\z2{\mathbb Z_2}

\begin{document}

\title{Quantum spin liquid on a triangular lattice with 7 particles}
\title{$\z2 \times \z2 \times \z2$ quantum spin liquid on a triangular lattice}
\title{Quantum spin liquid with 7 elementary particles in 2 dimensions}
\title{Quantum spin liquid with 7 elementary particles}

\author{Haoyu Wang}

\author{Hitesh J. Changlani}
\affiliation{Department of Physics and Astronomy,
	Johns Hopkins University,
	Baltimore, Maryland 21218, USA
}

\author{Yuan Wan}
\affiliation{Perimeter Institute for Theoretical Physics,
	Waterloo, Ontario N2L 2Y5 Canada}
	
\author{Oleg Tchernyshyov}
\affiliation{Department of Physics and Astronomy,
	Johns Hopkins University,
	Baltimore, Maryland 21218, USA
}

\begin{abstract}
We present an exactly solvable model of a quantum spin liquid with Abelian anyons in $d=2$ spatial dimensions. With spins 1/2 on a triangular lattice and six-body interactions, our model has zero spin correlation length and localized elementary excitations like the toric codes of Kitaev and Wen. In contrast to those earlier models, it has more elementary particles---4 bosons and 3 fermions---and higher topological degeneracy of 64 on a torus. Elementary excitations are boson-fermion pairs that come in 12 distinct flavors. We use string operators to expose the topological nature of the model.
\end{abstract}

\maketitle

\section{Introduction}

Quantum spin liquids (QSLs) have become a major focus point in magnetism research. Introduced in 1973 by Anderson as ``a fluid of mobile valence bonds'' \cite{MatResBull.8.153}, a QSL was initially seen as a result of a disruption of long-range magnetic order by strong quantum fluctuations. In $d=1$ spatial dimension, long-range order is disrupted in (almost) any magnet with a continuous global symmetry. Solvable models of QSLs in $d=1$ abound, including $S=1/2$ and $S=1$ antiferromagnetic chains \cite{ZPhys.71.205, PhysRevLett.59.799}, and so do their experimental realizations. It was later realized that, in the absence of a spontaneously broken continuous symmetry, elementary excitations in a QSL may be quite different from the familiar magnons of the ordered state and may carry fractional spin 1/2 \cite{PhysLettA.85.375}. Fractionalization of elementary excitations in a $S=1/2$ Heisenberg chain has been confirmed by inelastic neutron scattering, which transfers spin 1 to the magnet and creates two or more elementary excitations, thereby yielding a continuum of energies for a given momentum \cite{PhysRevLett.70.4003, NatPhys.9.435}.

In higher dimensions, $d \geq 2$, solvable models of QSLs are harder to come by. For a long time, clues about the nature of higher-dimensional QSLs came from approximate solutions based on slave-particle approaches, in which quasiparticles with fractional spin are smuggled in at the very beginning \cite{PhysRevB.37.3774, PhysRevLett.62.1694}. These approximate solutions brought an important insight that QSLs in higher dimensions may be closely related to lattice gauge theories. This connection revealed the presence of a subtle topological order in QSLs \cite{PhysRevB.44.2664} that arises from long-range quantum entanglement of spins and endows the elementary excitations of the system with anyon quantum statistics \cite{RepProgPhys.80.016502}.

These conjectures were confirmed with the advent of exactly solvable models of QSLs in $d=2$ dimensions. The first of those were spin-1/2 ``toric-code'' models on a square lattice of \textcite{Kitaev2003} and \textcite{Wen2003}. Kitaev's model is in essence a $\z2$ gauge theory; its elementary excitations come in the form of electric charges living on sites and magnetic fluxes of a $\z2$ lattice gauge field living on plaquettes. Closely related Wen's model also has two types of elementary excitations, which can be described as two distinct bosonic particles with mutual semion statistics. In both models, the creation of a single elementary excitation is a nonlocal process involving a physical transformation along a path extending to infinity (or to the edge of the system) implemented by a string operator. Owing to the highly entangled nature of the ground state, the exact path of a string is not important and can be deformed as long as the ends remain fixed. Simply put, the ends of a string are visible but the body is not.

A significant drawback of these models is the unnatural form of their Hamiltonians, which contain only four-spin interactions of a very specific form. These may be hard to realize in a magnetic material. (See, however, a recent proposal for realizing these models at an interface of a magnet and superconductor \cite{arXiv:1610.04614}.) This problem was later ameliorated in another solvable model by \textcite{Kitaev2006} with spins 1/2 on a honeycomb lattice. That model, on the one hand, is reducible to the toric-code models \textcite{Kitaev2003, Wen2003} in a certain limit and, on the other, has more realistic two-spin interactions. Its elementary excitations are magnetic fluxes of a $\z2$ lattice gauge field and Majorana fermions minimally coupled to the gauge field. \textcite{PhysRevLett.102.017205} pointed out a way of realizing the honeycomb model in magnets with strong spin-orbit coupling. Potential realizations in magnets with transition-metal ions are currently under experimental investigation \cite{NatMater.15.733}.

Our brief excursion into the history of quantum spin liquids was meant to underscore the important role of exactly solvable models. Even if a model does not seem realistic, it can provide a window into the realm of spin liquids. Further research may lead to the discovery of models with similar properties and better chances of being found in a real material. With this in mind, we present a new solvable model of a QSL in $d=2$ dimensions. As in Kitaev and Wen's models, the elementary particles are Abelian anyons. Its main difference from the models mentioned above is a larger number of elementary particles, which include 4 bosons and 3 fermions, all of them mutual semions with respect to one another. In Kitaev and Wen's models, the elementary particles are 2 bosons and 1 fermion. A larger number of elementary particles directly translates into higher topological degeneracy of energy eigenstates, 64 on a torus in our model versus 4 in Kitaev and Wen's QSLs.

In this paper, we distinguish between elementary \emph{excitations} and elementary \emph{particles}. We define the former as the smallest quanta of energy and the latter as the natural building blocks of the model. Although they happen to be the same in Kitaev and Wen's models, they need not be. In our model, there are 12 distinct types of lowest-energy excitations. However, they can be put together from a smaller number of basic building blocks. These building blocks themselves have a higher energy but they are conceptually simpler. One might point to a loose analogy with QCD, where quarks are elementary particles but not elementary excitations.

The paper is organized as follows. In Sec.~\ref{sec:model} we present the model and outline basic properties of its building blocks, closed and open strings. Sec.~\ref{sec:closed-strings} describes a systematic construction of closed strings, including non-contractible loops (e.g., on a torus), whose algebra determines the topological degeneracy. In Sec.~\ref{sec:open-strings} we construct open strings, characterize elementary particles living at their ends, and build elementary excitations of the Hamiltonian out of them. Edge states are discussed in Sec.~\ref{sec:edge}. In Sec.~\ref{sec:numerics} we present numerically obtained energy spectra of small clusters to corroborate our theoretical analysis. We summarize our results in Sec.~\ref{sec:discussion}.

\section{The model and its building blocks}
\label{sec:model}

\subsection{The Hamiltonian}

\begin{figure}
\includegraphics[width=0.95\columnwidth]{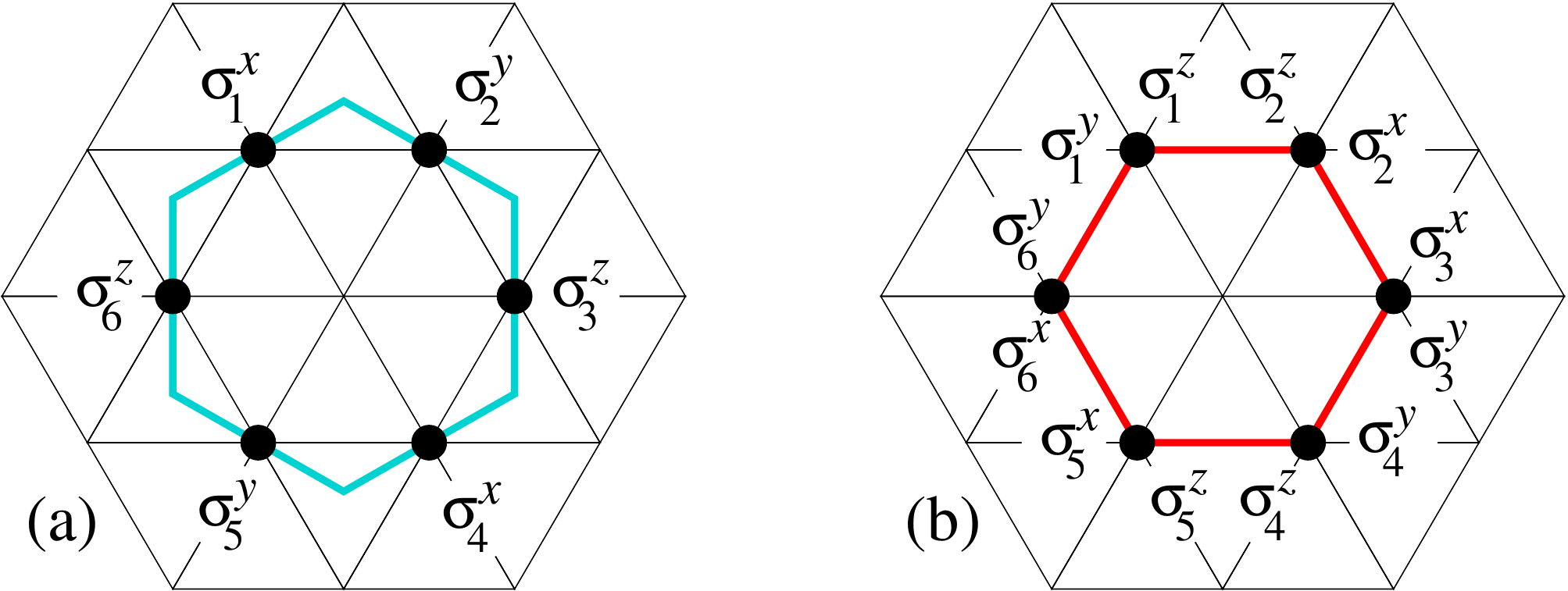}
\caption{(a) The 6-body spin interaction (\ref{eq:W}). (b) Its alternative representation (\ref{eq:W-alt}).}
\label{fig:W}
\end{figure}

Our model has spins 1/2 residing on sites of a triangular lattice. The Hamiltonian is a sum of six-spin interactions,
\begin{equation}
H = - \sum_{n} W_{n},
\label{eq:H-W}
\end{equation}
where the six-spin operator
\begin{equation}
W_n =  \sigma_6^z \sigma_5^y \sigma_4^x \sigma_3^z \sigma_2^y \sigma_1^x,
\label{eq:W}
\end{equation}
is borrowed from the honeycomb model of \textcite{Kitaev2006}; sites $1$ through $6$ are the nearest neighbors of site $n$, Fig. \ref{fig:W}(a).  It has an alternative representation
\begin{equation}
W_n =
	(\sigma_1^y\sigma_6^y)	
	(\sigma_6^x\sigma_5^x)
	(\sigma_5^z\sigma_4^z)
	(\sigma_4^y\sigma_3^y)
	(\sigma_3^x\sigma_2^x)
	(\sigma_2^z\sigma_1^z),
\label{eq:W-alt}
\end{equation}
Fig.~\ref{fig:W}(b), in which the flavor of the Pauli matrices on a link is determined by its orientation \cite{Kitaev2006, Petrova2014}.

As in the honeycomb model, all $W_n$ operators commute with one another and thus can be simultaneously diagonalized. The number of hexagons on a triangular lattice is the same as the number of sites as every hexagon is centered on a site. Thus the $W_n$ operators represent a complete set of observables that can be used to specify the quantum state of the system. (The usual caveats apply. E.g., the presence of an edge reduces the number of hexagons relative to that of sites, leading to higher degeneracy.)

The state of lowest energy is achieved when we set $W_n=+1$ for every hexagon. Elementary excitations are hexagons with $W_n=-1$. The energetics are reminiscent of the toric-code models on a square lattice of Kitaev \cite{Kitaev2003} and Wen \cite{Wen2003}. However, we shall see that the nature of elementary excitations and the spectrum of elementary particles are different in our model.

\begin{figure}
\includegraphics[width=0.9\columnwidth]{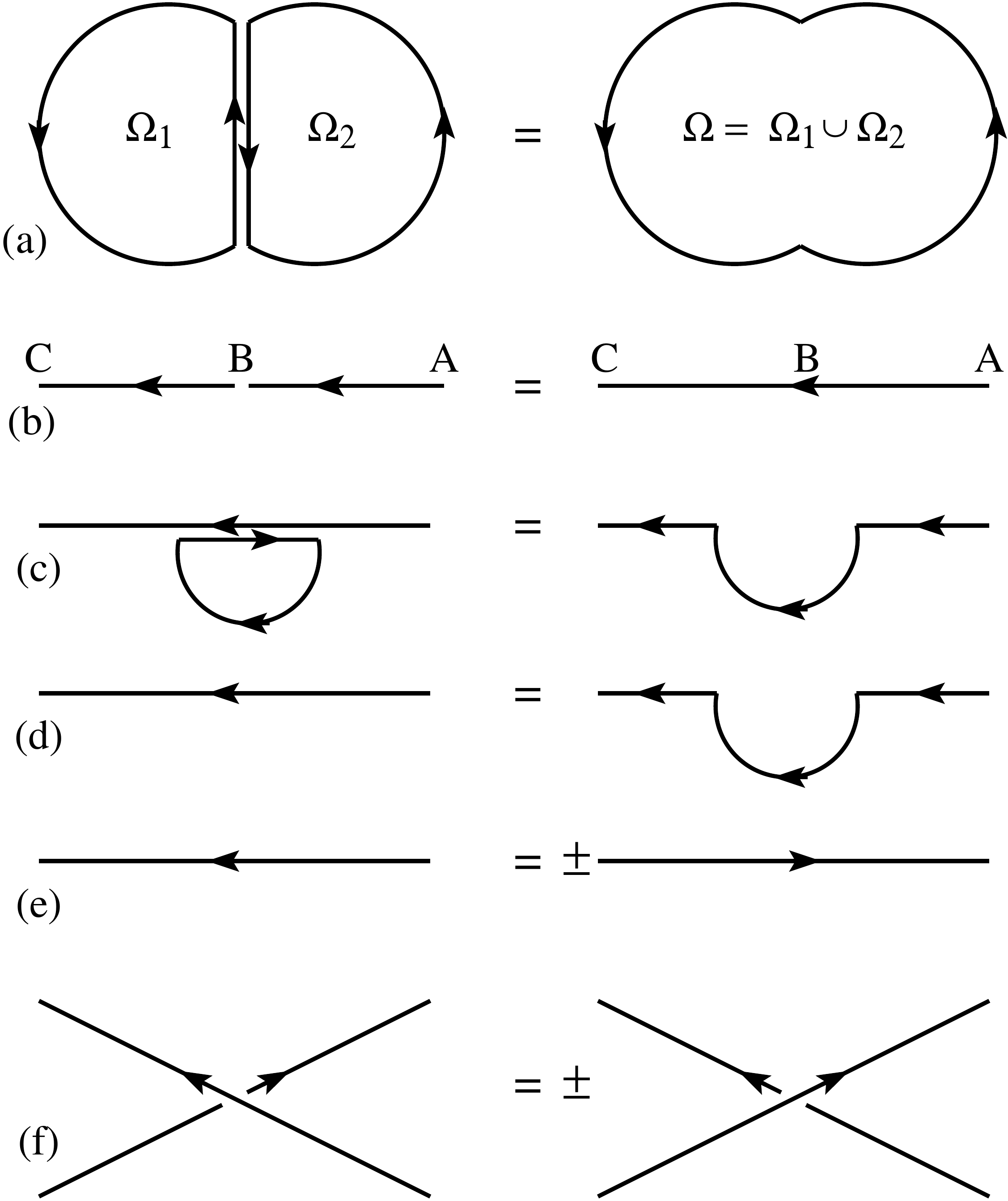}
\caption{Basic properties of strings. (a) Multiplication of two closed strings sharing a segment. (b) Concatenation of two open strings sharing a point. (c) Deformation of a string by the attachment of a closed loop.  (d) In a ground state, a small deformation does not change the value of a string. (e) Reversal of the string direction may change its sign. (f)} Strings intersecting once either commute or anticommute.
\label{fig:strings-rules}
\end{figure}

\subsection{Basic properties of strings}

The basic building block in our model is the string $W_{\mathrm T}$, a product of spin-1/2 Pauli operators along some lattice path $\mathrm T$. Non-commutativity of Pauli operators means that the value of a string operator may depend on its direction. Strings can be closed or open. The simplest examples of closed strings are operators $W_n$, Fig.~\ref{fig:W}. We will use them to construct longer loops in Sec.~\ref{sec:closed-strings} and open strings in Sec.~\ref{sec:open-strings}. Here we briefly survey general properties of string operators. For simplicity, we abstract from lattice details and depict strings as continuous lines in Fig.~\ref{fig:strings-rules}.

\begin{enumerate}[(a)]

\item A closed string may follow a path $\mathrm T = \partial \Omega$ around the boundary of a region $\Omega$. Consider two adjacent but non-overlapping regions $\Omega_1$ and $\Omega_2$. The product of two closed strings $W_{\partial \Omega_1}$ and $W_{\partial \Omega_2}$ yields a string living on the boundary of the combined region $\Omega = \Omega_1 \cup \Omega_2$, Fig.~\ref{fig:strings-rules}(a):
\begin{equation}
W_{\partial \Omega_1} W_{\partial \Omega_2} = W_{\partial \Omega}.
\end{equation}

\item Two open strings sharing a point can be concatenated, Fig.~\ref{fig:strings-rules}(b):
\begin{equation}
W_{CB} W_{BA} = W_{CBA}.
\end{equation}

\item A string can be deformed by attaching to it a closed loop, Fig.~\ref{fig:strings-rules}(c).

\item In a ground state of our model, a small deformation of a string---the attachment of a hexagon---does not change its value, Fig.~\ref{fig:strings-rules}(d).

\item Strings are generally directional, Fig.~\ref{fig:strings-rules}(e). A reversed string differs from the original by a factor of $\pm 1$, depending on its type and length.

\item Two strings intersecting once either commute or anticommute,
\begin{equation}
W_1 W_2 = \pm W_2 W_1,
\end{equation}
depending on their types, Fig.~\ref{fig:strings-rules}(f). Two strings with an even number of intersections commute. Thus any two contractible loops commute. An example of that is the commutativity of all $W_n$ operators (\ref{eq:W}).

\end{enumerate}

Note that strings are generally directed, so operators $W_{AB}$ and $W_{BA}$ are not necessarily identical. In our model, $W_{AB} = \pm W_{BA}$, where the sign depends on the type of the string and on its length.

\section{Closed strings.}
\label{sec:closed-strings}

\subsection{Fermionic Wilson loops}

\begin{figure}
\includegraphics[width=0.95\columnwidth]{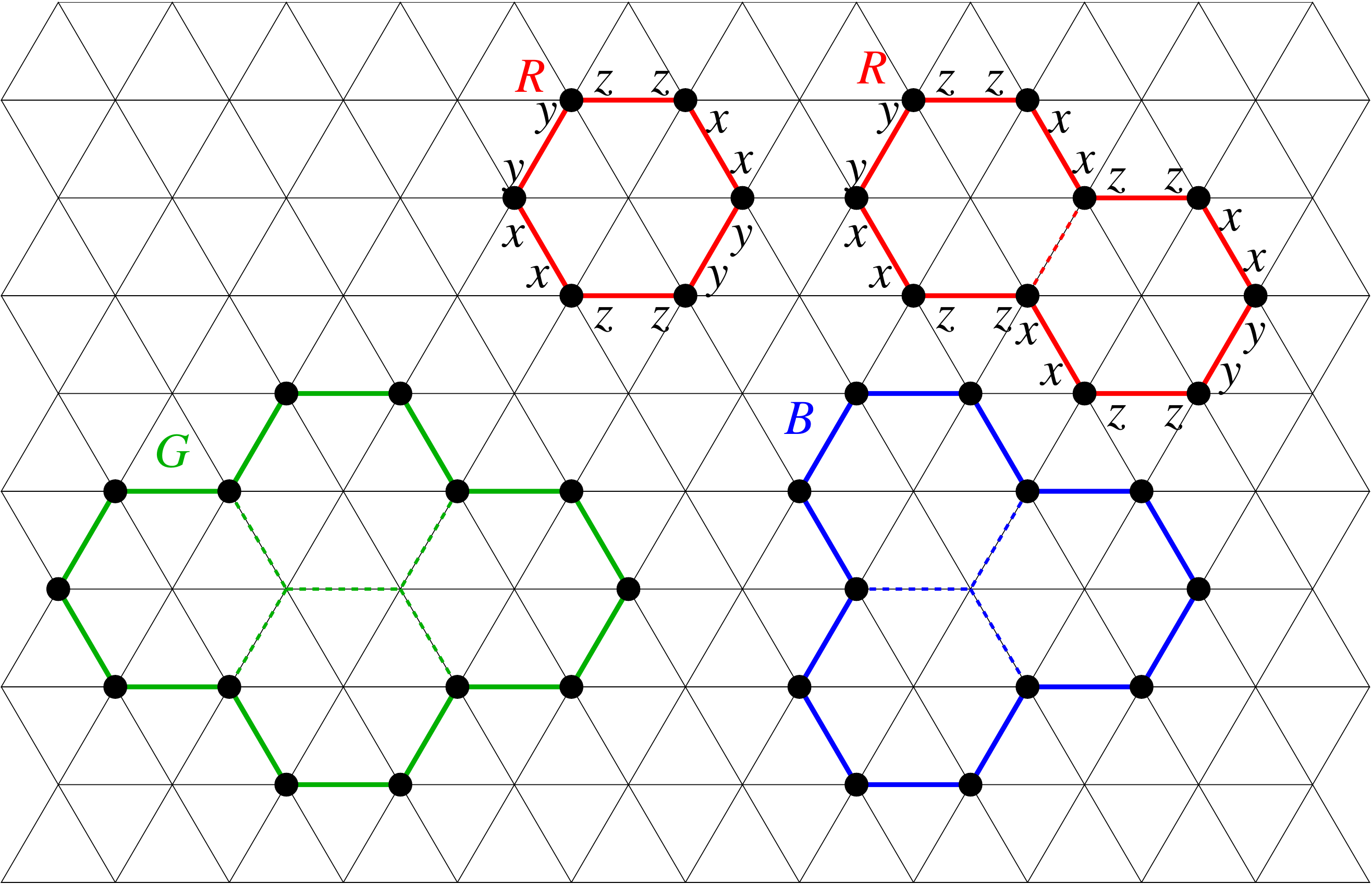}
\caption{Fermionic Wilson loops live on honeycombs. Loops of the same color can be deformed into one another by attaching or removing hexagons. }
\label{fig:F-loops}
\end{figure}

To gain further insight into the physics of the model, it is helpful to view the operator $W_n$ as a Wilson loop measuring the $\z2$ magnetic flux through a hexagon \cite{Kitaev2003}. To use an analogy with a more familiar $U(1)$ gauge theory, a particle with electric charge $e$ moving around the boundary $\partial \Omega$ of some region $\Omega$ picks up the Aharonov-Bohm phase
\begin{equation}
\phi = \frac{e}{\hbar c} \oint_{\partial \Omega} \mathbf A \cdot d \mathbf r
= \frac{e}{\hbar c} \int_\Omega \mathbf B \cdot d^2 \mathbf r
= \frac{e \Phi_\Omega}{\hbar c}
\end{equation}
proportional to the magnetic flux $\Phi_\Omega$ through the area. In this analogy, $\Omega$ is a hexagon, $\partial \Omega$ is its perimeter, and $W = e^{i \phi}$, where $\phi = 0$ or $\pi$ in a $\z2$ gauge theory. Whereas the Aharonov-Bohm phase $\phi$ is additive, its exponential $W = e^{i \phi}$ is multiplicative. It can be checked with the aid of representation (\ref{eq:W-alt}) that a product of $W$ operators for two or more edge-sharing hexagons forming a cluster $\Omega$ yields a Wilson loop along the cluster boundary $\partial \Omega$ (Fig.~\ref{fig:F-loops}):
\begin{equation}
W_{\partial \Omega} = \prod_{n \in \Omega} W_n
= 	(\sigma_n^\nu \sigma_{n-1}^\nu)
	\ldots
	(\sigma_2^\beta \sigma_1^\beta)
	(\sigma_1^\alpha \sigma_n^\alpha)
\label{eq:F-P}
\end{equation}
The flavor $\alpha = x, y, z$ of the Pauli operators in a link operator $\sigma_1^\alpha \sigma_n^\alpha$ depends on the orientation of the link in the same way as for a hexagon, Eq.~(\ref{eq:W-alt}) and Fig.~\ref{fig:F-loops}.

\begin{figure}
\includegraphics[width=0.95\columnwidth]{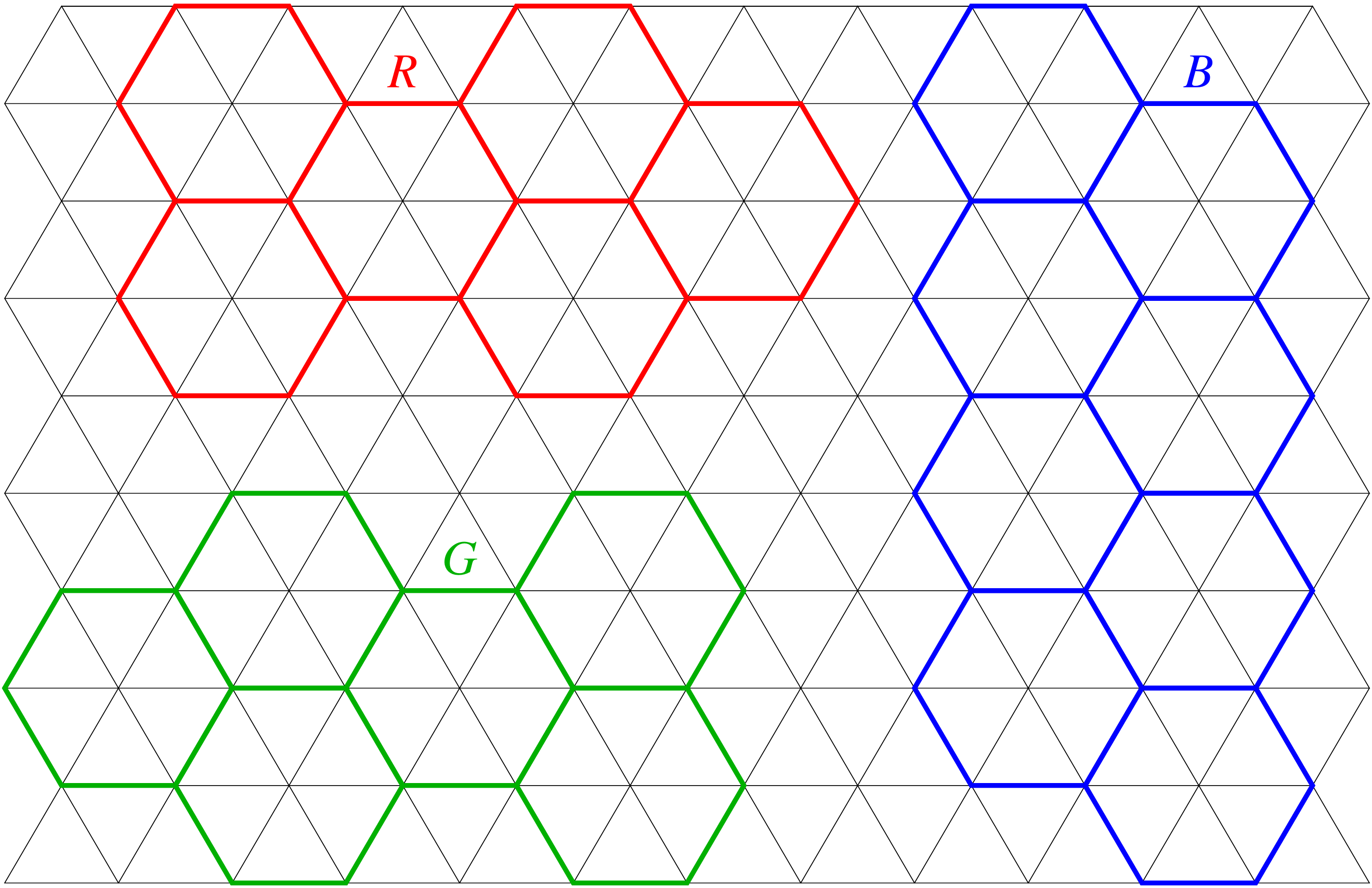}
\caption{Three honeycombs---red (R), green (G), and blue (B)---with links on the original triangular lattice.}
\label{fig:F-honeycombs}
\end{figure}

A closed string constructed in this way lives on links of a honeycomb lattice obtained by removing every third site of the original triangular lattice. This task can be accomplished in three ways, with different sublattices removed, Fig.~\ref{fig:F-honeycombs}. The three resulting honeycombs do not share links. Strings living on different honeycombs cannot be deformed into one another by elementary deformations consisting of attaching or removing a hexagon. We thus find three distinct types of loops labeled red ($R$), green ($G$), and blue ($B$).

As we shall see later (Sec.~\ref{sec:open-strings}), Wilson loops of this type are associated with fermionic particles. We shall therefore refer to them as fermionic Wilson loops.

\begin{figure}
\includegraphics[width=0.95\columnwidth]{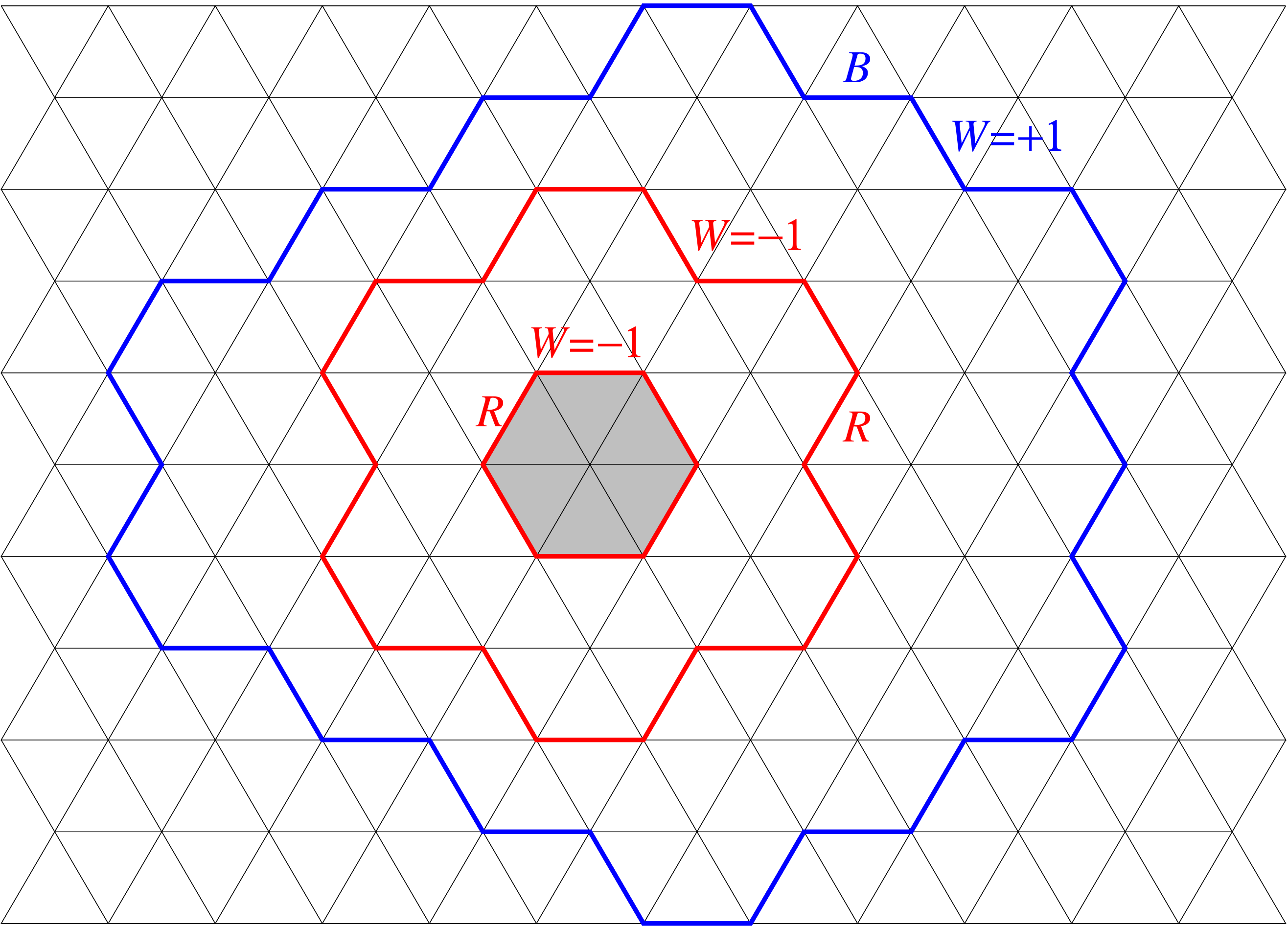}
\caption{The shaded hexagon contains an elementary excitation $W = -1$. The little red loop on its perimeter and the big red loop on the same honeycomb return a nontrivial Aharonov-Bohm phase $W = -1$. The blue loop belongs to a different honeycomb and thus has $W = +1$.}
\label{fig:F-loops-2}
\end{figure}

The three types of fermionic Wilson loops provide independent physical information. Suppose there is a single excited plaquette ($W_n = -1$) in the middle of a large area, Fig.~\ref{fig:F-loops-2}. Its presence can be detected by measuring the value of a Wilson loop $F$ enclosing it. Only one type of Wilson loop---in this case, red---will have the nontrivial value $W=-1$, the other two will have $W=+1$.

Construction of the fermionic Wilson loop (\ref{eq:F-P}) with Pauli operators on the boundary of a cluster but not in its bulk was enabled by the following property of elementary Wilson loops (\ref{eq:W}). A site at the intersection of three hexagons contributes operators $\sigma^x$, $\sigma^y$, and $\sigma^z$ to their Wilson loops. It thus contributes $\sigma^x \sigma^y \sigma^z = i$, a number, to the product of $W_n$ operators of the cluster.

Our construction of fermionic Wilson loops closely matches that of Kitaev for his honeycomb model \cite{Kitaev2003}. The main difference is that our triangular lattice contains three honeycomb lattices, Fig.~\ref{fig:F-honeycombs}, which leads to the existence of three distinct flavors of fermionic strings instead of one. That, in turn, will give rise to higher topological degeneracy, as will be seen in Sec.~\ref{sec:closed-loops-global}.

\subsection{Bosonic Wilson loops}

Another way to construct big Wilson loops is to make a product of $W$ for elementary hexagons sharing corners, rather than edges. Two hexagons sharing a corner both contribute the same operator $\sigma^\alpha$. Again, a product of $W_n$ operators over a cluster of corner-sharing hexagons will only have Pauli operators on the boundary but not in the bulk of the cluster. This observation allows us to construct four more distinct types of strings.

\begin{figure}
\includegraphics[width=0.95\columnwidth]{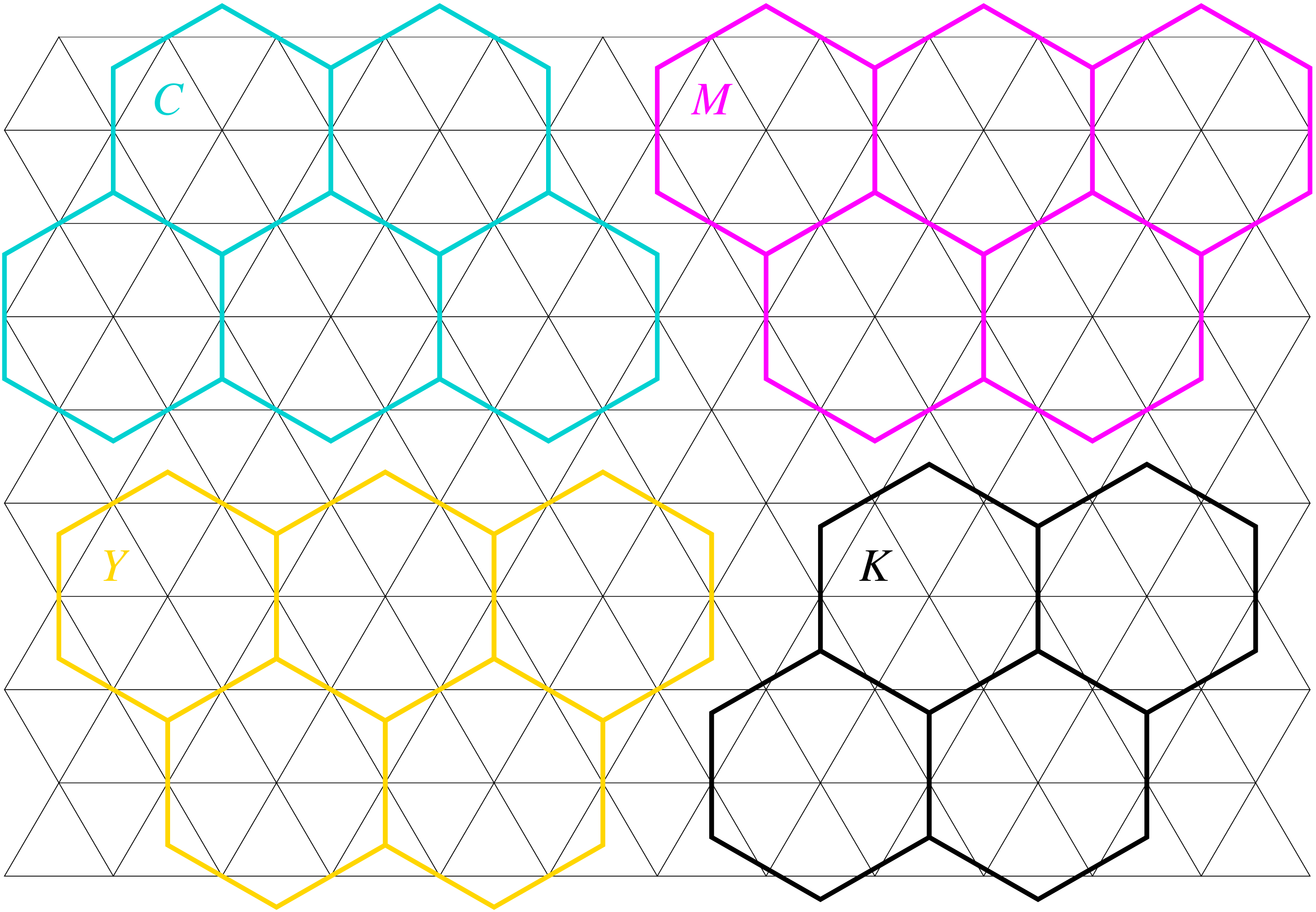}
\caption{Four dual honeycombs---cyan (C), magenta (M), yellow (Y), and black (K).}
\label{fig:B-honeycombs}
\end{figure}

\begin{figure}
\includegraphics[width=0.95\columnwidth]{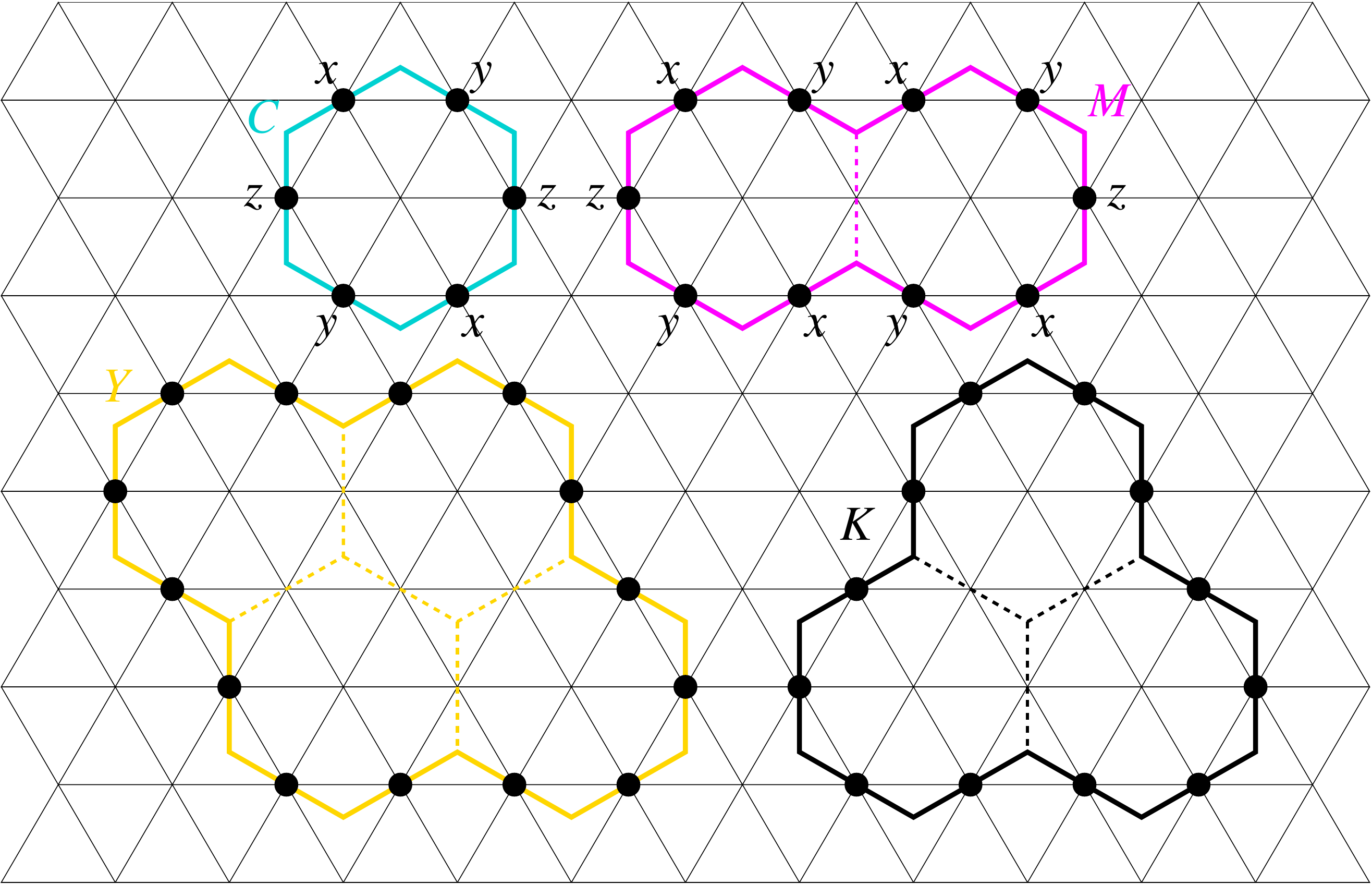}
\caption{Bosonic Wilson loops on dual honeycombs.}
\label{fig:B-loops}
\end{figure}

To facilitate this construction, it is convenient to introduce dual honeycombs whose links connect centers of triangles of the original lattice, Fig.~\ref{fig:B-honeycombs}. There are four distinct dual honeycombs labeled cyan ($C$), magenta ($M$), yellow ($Y$), and black ($K$).

An elementary hexagon can be identified with the shortest loop on a dual honeycomb, Fig.~\ref{fig:B-loops}. Longer dual loops can be constructed from products of $W$ operators on the same dual honeycomb:
\begin{equation}
W_{\partial \omega}
	= \prod_{n \in \omega} W_n
 	= \sigma_n^\nu \ldots \sigma_2^\beta \sigma_1^\alpha.
\label{eq:B-P}
\end{equation}
Here $\omega$ is a cluster of dual hexagons residing on the same dual honeycomb and $\partial \omega$ is its perimeter containing points $1, 2, \ldots, n$. The flavor $\alpha$ of the Pauli operator $\sigma_n^\alpha$ is determined by the orientation of the dual link passing through site $n$, Fig.~\ref{fig:B-loops}.

As we shall see later, strings on dual honeycombs are associated with bosonic particles, so we will refer to them as bosonic.

The existence of multiple string types---3 fermionic and 4 bosonic---has an analog in Wen's model \cite{Wen2003}, which has 1 fermionic and 2 bosonic strings \cite{PhysRevD.68.065003}.

\subsection{Strings are invisible}

Big Wilson loops $W$ commute with the Hamiltonian and are thus integrals of motion. However, they do not carry additional information beyond that which is already contained in elementary Wilson loops $W_n$ (\ref{eq:W}). Thus, in a ground state $|\psi_0\rangle$, all Wilson loops constructed out of short loops $W$ are trivial,
\begin{equation}
W|\psi_0\rangle = |\psi_0\rangle.
\end{equation}

A simple way to phrase this result is to say that closed strings are invisible in a ground state. Indeed, we can deform any loop (on an original or dual honeycomb) by attaching or removing hexagons to it. Thus the action of a $W$ operator on a ground state is invariant under deformations of the Wilson loop. If a Wilson loop is contractible to a point, its value must be trivial, $W = +1$.

\subsection{Global loops}
\label{sec:closed-loops-global}

A system with a nontrivial topology (e.g., a torus) will have non-contractible loops $\mathrm{T}$ on the original lattice or $\tau$ on the dual one that wind around the system. We can use the same prescriptions (\ref{eq:F-P}) and (\ref{eq:B-P}) to construct such non-contractible loops $S_\mathrm{T}$ and $S_\tau$. These global loop operators commute with short loops $W$ and thus with the Hamiltonian (\ref{eq:H-W}). However, because they are not reducible to a product of $W$ operators, their eigenvalues in a ground state are not necessarily $+1$. Thus we can use them to determine the topological degeneracy.

\begin{figure}
\includegraphics[width=0.95\columnwidth]{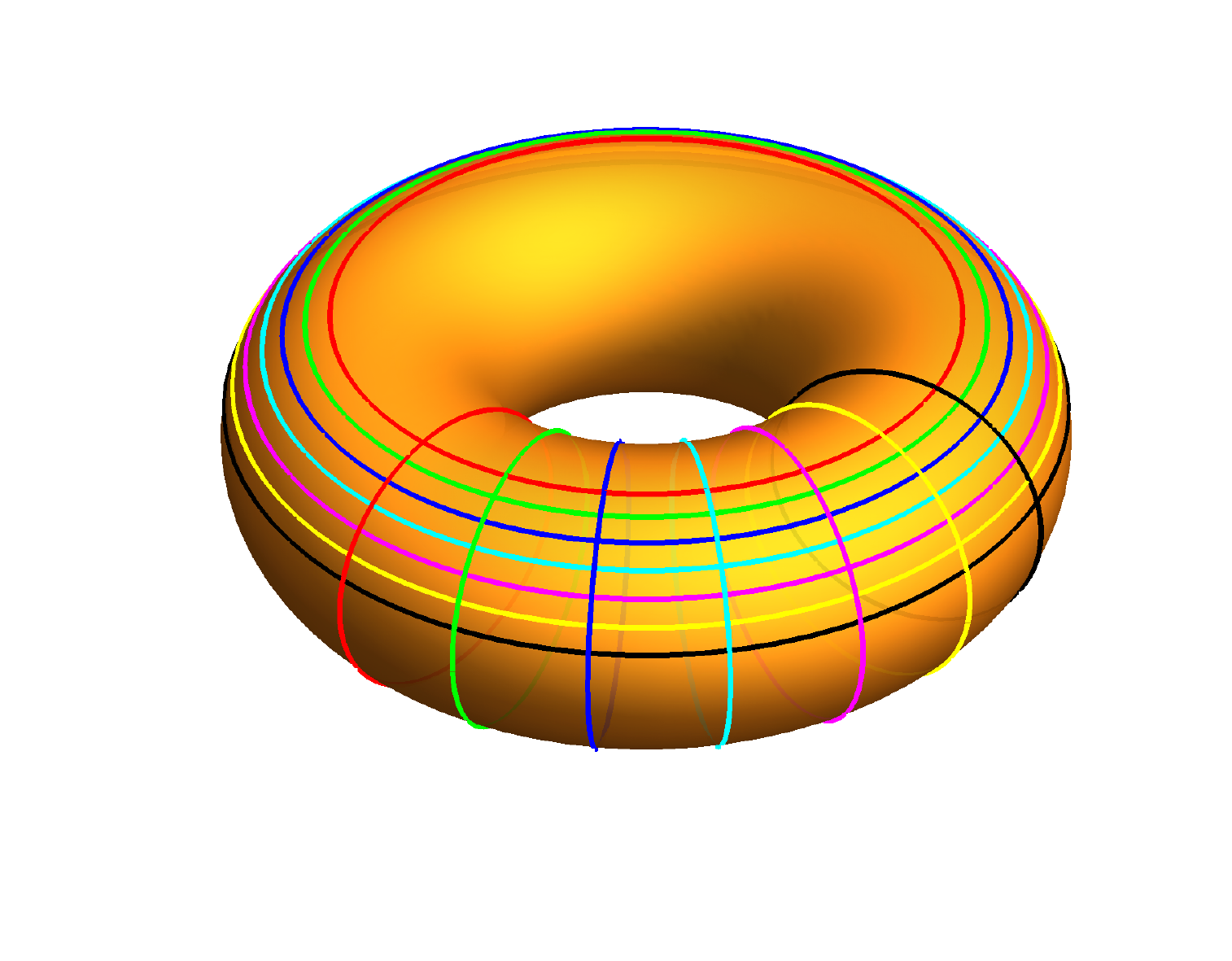}
\caption{Fourteen topologically distinct global loops winding in the two directions of the torus.}
\label{fig:global-loops}
\end{figure}

On a torus, viewed as a rectangle with periodic boundary conditions along both $x$ and $y$ directions, we obtain 7 topologically distinct loops winding around the lattice in the $x$ direction (Fig.~\ref{fig:global-loops}): 3 fermionic $W_{Rx}$, $W_{Gx}$, $W_{Bx}$ and 4 bosonic $W_{Cx}$, $W_{Mx}$, $W_{Yx}$, $W_{Kx}$;  7 more wind around the torus in the $y$ direction.

\begin{figure}
\includegraphics[width=0.95\columnwidth]{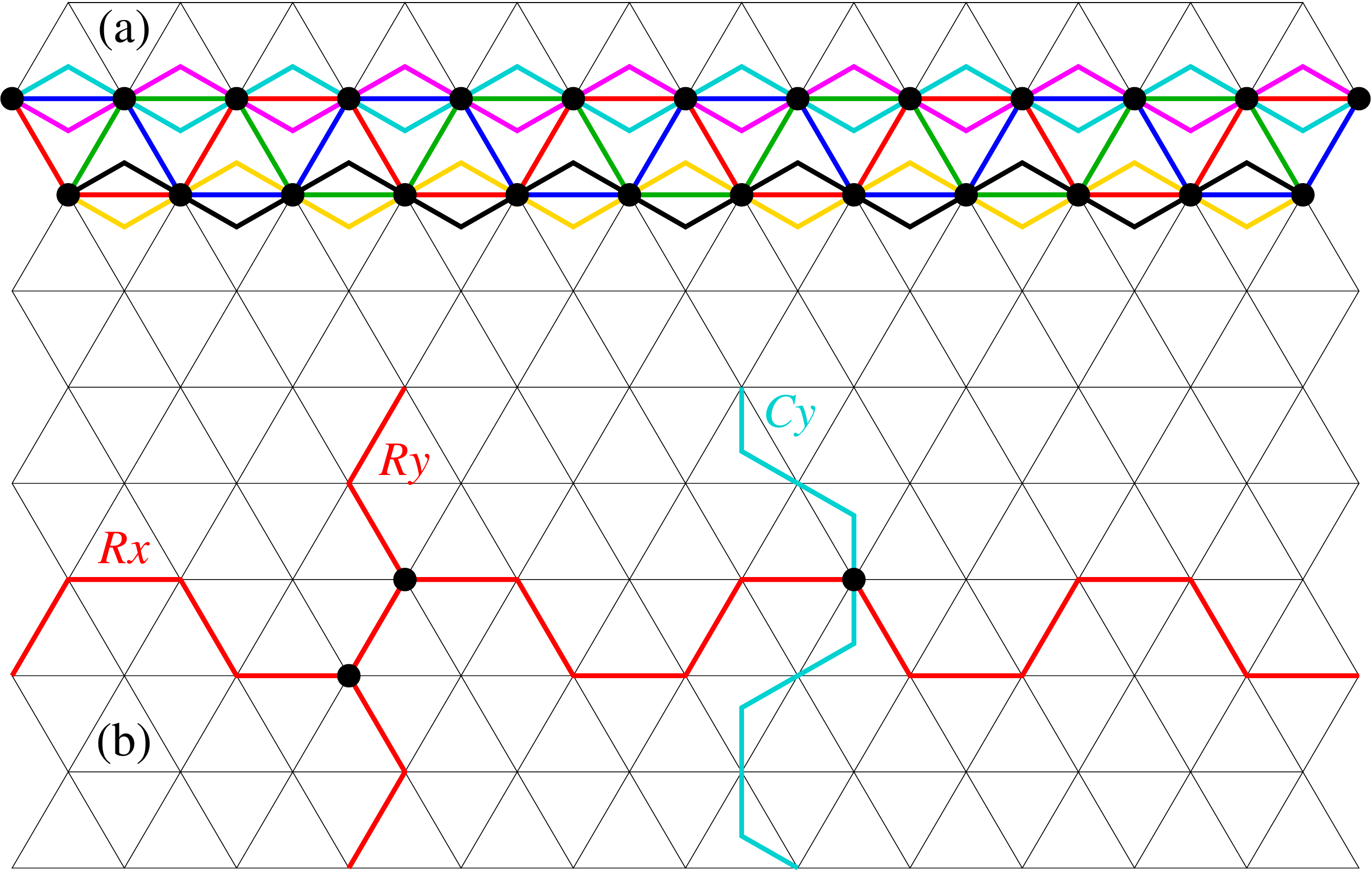}
\caption{(a) Seven overlapping strings cancel out as at each intersection we have a product of two operators $\sigma^x$ and two operators $\sigma^y$. (b) Global loops $W_{Rx}$ and $W_{Ry}$ commute, whereas $W_{Rx}$ and $W_{Ky}$ anticommute. Note that strings $R_x$ and $R_y$ are considered to have a single intersection, even though they share two lattice sites.}
\label{fig:constraint-commutation}
\end{figure}

Two global Wilson loops winding around the torus in the same direction commute because we can deform them to avoid any intersections, Fig.~\ref{fig:global-loops}. We can thus use their eigenvalues to label ground states:
\begin{equation}
|\psi_0\rangle = |W_{Rx}, W_{Gx}, W_{Bx}, W_{Cx}, W_{Mx}, W_{Yx}, W_{Kx} \rangle.
\label{eq:ground-states-torus}
\end{equation}
If these operators could independently take on eigenvalues $\pm 1$, we would obtain $2^7$ ground states on a torus. However, they are not fully independent and satisfy the constraint
\begin{equation}
W_{Rx} W_{Gx} W_{Bx} W_{Cx} W_{Mx} W_{Yx} W_{Kx} = +1,
\label{eq:constraint}
\end{equation}
which can be checked by examining a segment of 7 overlapping strings, where all Pauli operators cancel out, Fig.~\ref{fig:constraint-commutation}(a). The constraint (\ref{eq:constraint}) reduces the number of possible ground states (\ref{eq:ground-states-torus}) to $2^{7-1} = 64$. We may use the first six bits in Eq.~(\ref{eq:ground-states-torus}) to designate a ground state; the seventh, $W_{Kx}$, adjusts as necessary to satisfy the parity constraint (\ref{eq:constraint}).

As can be seen in Fig.~\ref{fig:global-loops}, two global loops winding along different directions of a torus intersect once (or, more generally, an odd number of times). The two Wilson operators commute if the loops are of the same type (e.g., both $R$ or both $M$) and anticommute otherwise, Fig.~\ref{fig:constraint-commutation}(b). Thus we may use Wilson operator $W_{Ry}$ to alter all bits in a ground state (\ref{eq:ground-states-torus}) except for $W_{Rx}$. Even more convenient would be to use a combination such as $W_{Ry} W_{Ky}$, which alters the encoding bit $W_{Rx}$ as well as the ancillary $W_{Kx}$. In this way, we can access all $2^6$ degenerate ground states starting from any of them.

\section{Open strings}
\label{sec:open-strings}

\subsection{Open fermionic strings}

\begin{figure}
\includegraphics[width=0.95\columnwidth]{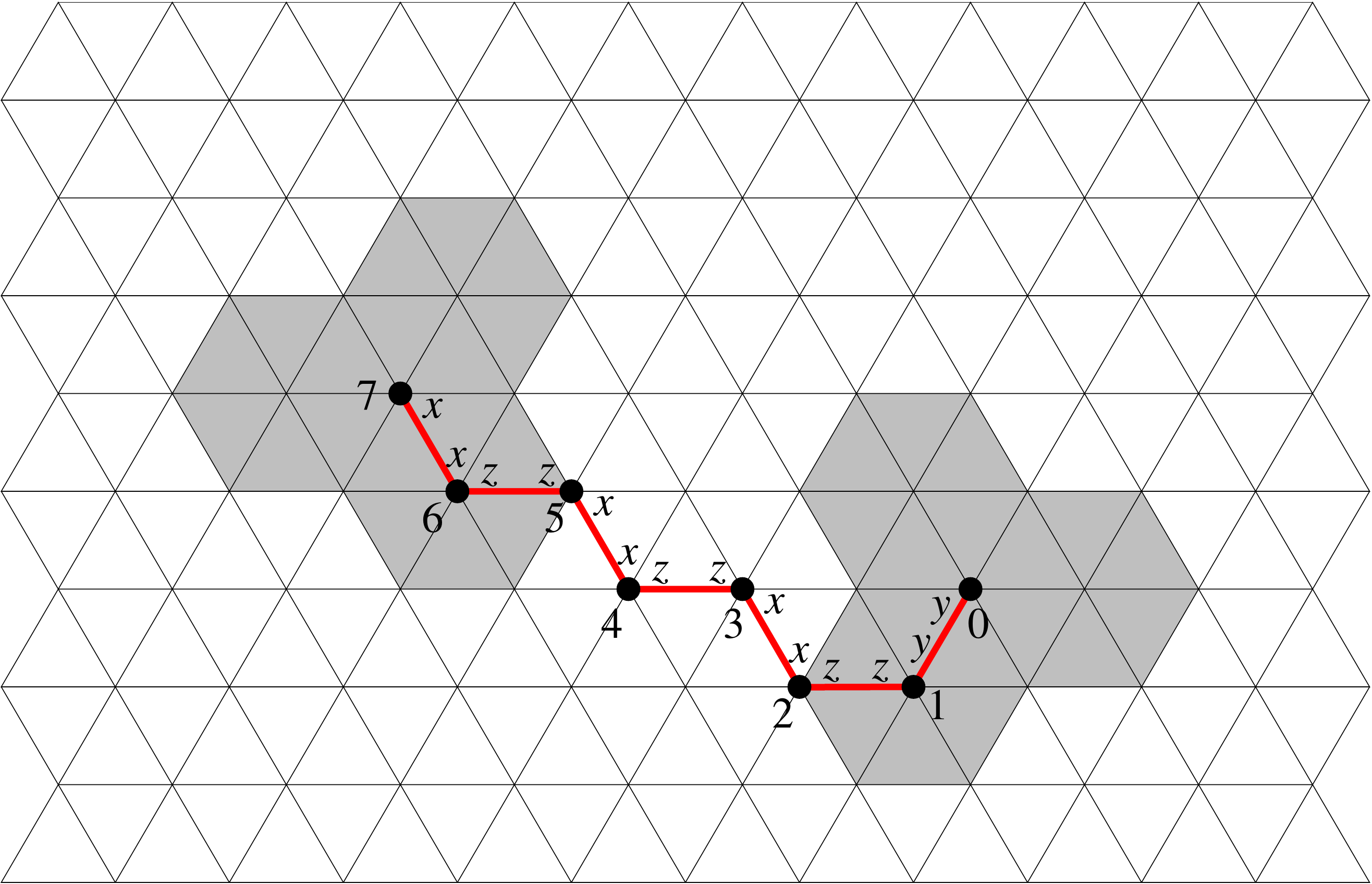}
\caption{Open fermionic string $S_{\mathrm T} = (\sigma_7^x \sigma_6^x) \ldots (\sigma_1^y \sigma_0^y)$ creates excitations within the shaded areas.}
\label{fig:F-string}
\end{figure}

Another way to make nontrivial string operators is to cut a loop and make a string with open ends:
\begin{equation}
S_\mathrm{T} =
	(\sigma_n^\nu \sigma_{n-1}^\nu)
	\ldots
	(\sigma_2^\beta \sigma_1^\beta)
	(\sigma_1^\alpha \sigma_0^\alpha),
\label{eq:open-F-string-def}
\end{equation}
where $\mathrm{T}$ is a path on a honeycomb lattice traversing points $0, 1, \ldots, n$, Fig.~\ref{fig:F-string}. The order in which the path $\mathrm{T}$ is traveled matters:
\begin{equation}
S_{\mathrm{T}^{-1}} = (-1)^{|\mathrm{T}|+1} S_\mathrm{T},
\end{equation}
where the path length $|\mathrm T|$ equals the number of links.

When $\mathrm T$ is a simple loop (every site is traversed once), its string operator (\ref{eq:open-F-string-def}) reduces to a Wilson loop (\ref{eq:F-P}).

Open strings (\ref{eq:open-F-string-def}) are multiplicative under concatenation. If the tail of path $\mathrm T_1$ coincides with the head of path $\mathrm T_2$ then
\begin{equation}
S_{\mathrm{T}_2} S_{\mathrm{T}_1} = S_{\mathrm{T}_2 \mathrm{T}_1}.
\label{eq:concatenation-F}
\end{equation}

\begin{figure}
\includegraphics[width=0.95\columnwidth]{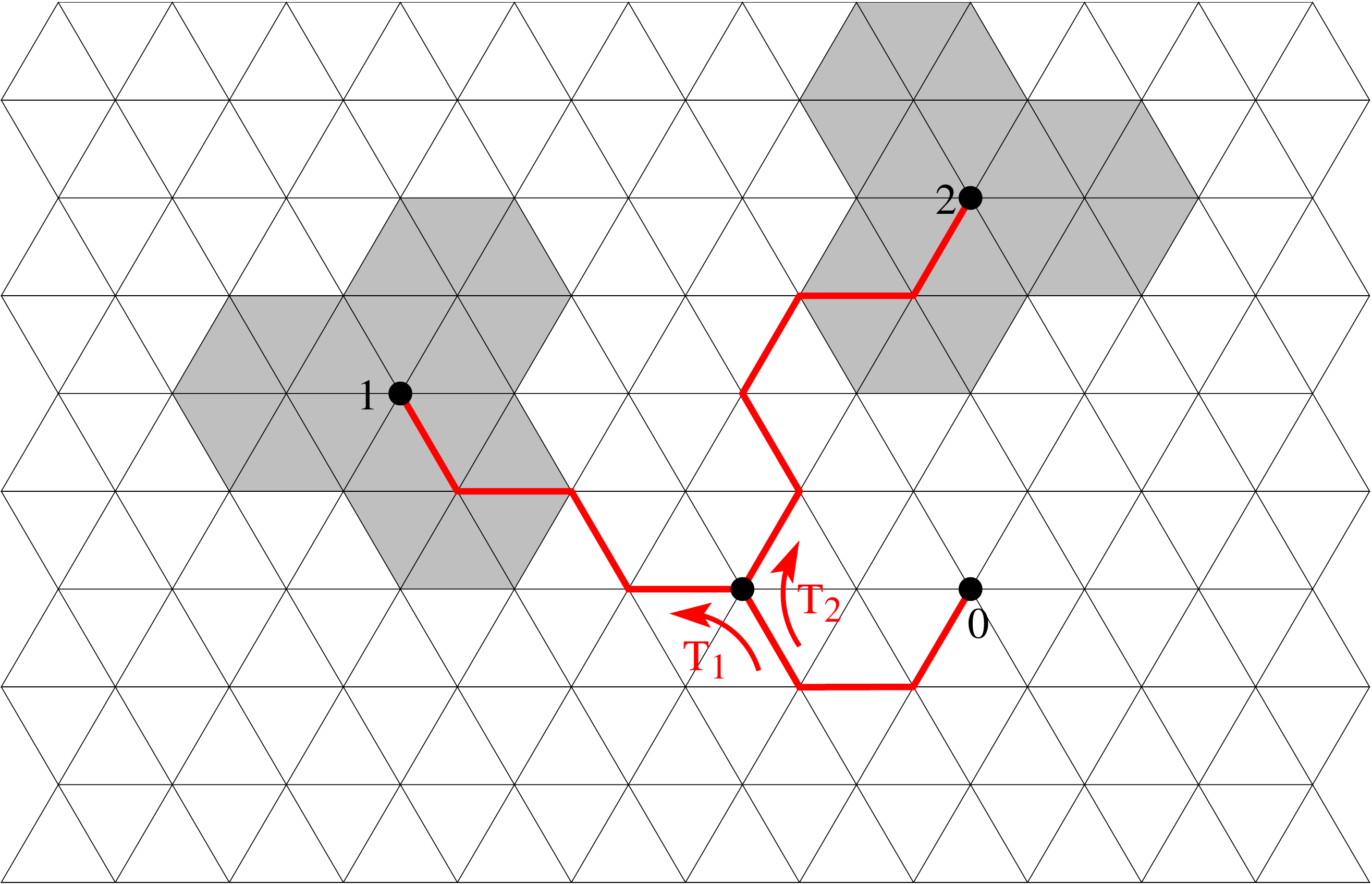}
\caption{Two string operators $S_{\mathrm T_1}$ and $S_{\mathrm T_2}$ with heads at 0 and tails at 1 and 2, respectively, create excitations near the tails (shaded areas). }
\label{fig:F-strings-anticommute}
\end{figure}

When two paths overlap and have a single branching point, their string operators anti-commute. In Fig.~\ref{fig:F-strings-anticommute}, strings $\mathrm T_1$ and $\mathrm T_2$ connect point 0 with points 1 and 2, respectively. The product of string operators
\begin{equation}
S_{\mathrm T_1} S_{\mathrm T_2} = - S_{\mathrm T_2} S_{\mathrm T_1}
\end{equation}
creates excitations near the end points 1 and 2. The anti-commutation of $S_{\mathrm T_1}$ and $S_{\mathrm T_2}$ implies that the resulting excitations are fermions. See Sec.~\ref{sec:open-strings-braiding}.

The fermionic nature of strings $S_{\mathrm T}$ becomes manifest if one uses Kitaev's representation of spin operators in terms of Majorana fermions \cite{Kitaev2006}, $\sigma_n^\alpha = i b_n^\alpha c_n$. Eq.~(\ref{eq:open-F-string-def}) translates into a product of $c$ Majorana operators at the ends connected by a $\z2$ gauge string,
\begin{equation}
S_{\mathrm T} = c_n (-i u_{n,n-1}) \ldots (-i u_{21})(-i u_{10}) c_0.
\end{equation}
It should be noted that link operators $u_{mn} = i b_m^\alpha b_n^\alpha = - u_{nm}$ commute among themselves only for a single honeycomb (as is the case in Kitaev's honeycomb model). Link operators belonging to different honeycombs anticommute if the two links share a site and point in the same direction.

\subsection{Open bosonic strings}

\begin{figure}
\includegraphics[width=0.95\columnwidth]{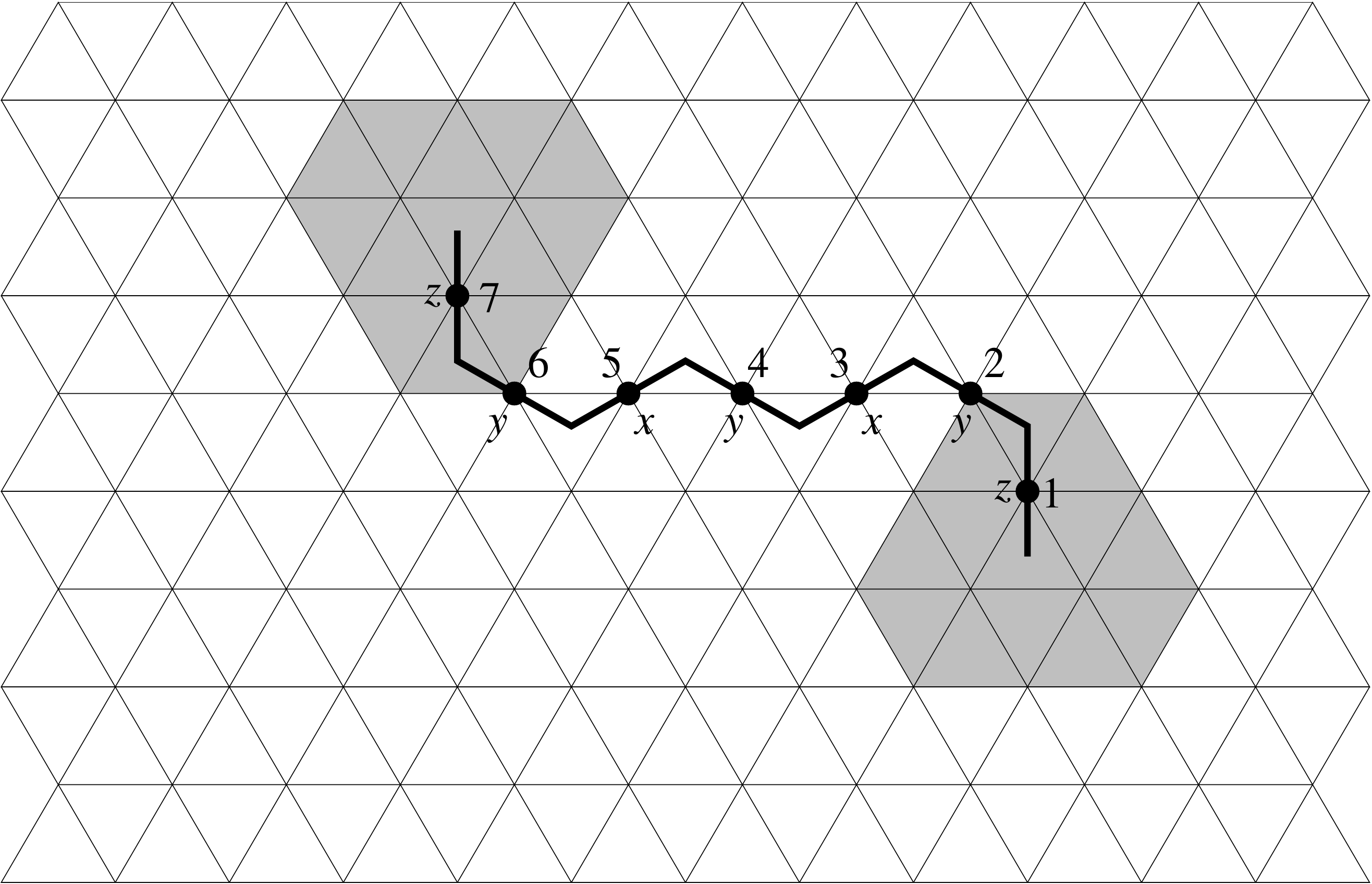}
\caption{An open bosonic string $S_{\tau} = \sigma_7^z \sigma_6^y \ldots \sigma_2^y \sigma_1^z$ creates excitations within the shaded areas.}
\label{fig:B-string}
\end{figure}

In a similar way, we define a dual open string,
\begin{equation}
S_{\tau} =  \sigma_n^\nu \ldots \sigma_2^\beta \sigma_1^\alpha,
\label{eq:open-B-string-def}
\end{equation}
where $\tau$ is a path on a dual honeycomb connecting sites $1, 2, \ldots, n$, Fig.~\ref{fig:B-string}. Dual strings are not sensitive to the direction of traverse:
\begin{equation}
S_{\tau^{-1}} = S_\tau.
\end{equation}
Strings on the same dual honeycomb can be concatenated,
\begin{equation}
S_{\tau_2} S_{\tau_1} = S_{\tau_2 \tau_1},
\end{equation}
and commute with one another,
\begin{equation}
S_{\tau_2} S_{\tau_1} = S_{\tau_1} S_{\tau_2}.
\end{equation}
Thus excitations at the ends of dual open strings are bosons.

\subsection{Braiding statistics}
\label{sec:open-strings-braiding}

\begin{widetext}

\begin{figure}[t]
\includegraphics[width=0.32\columnwidth]{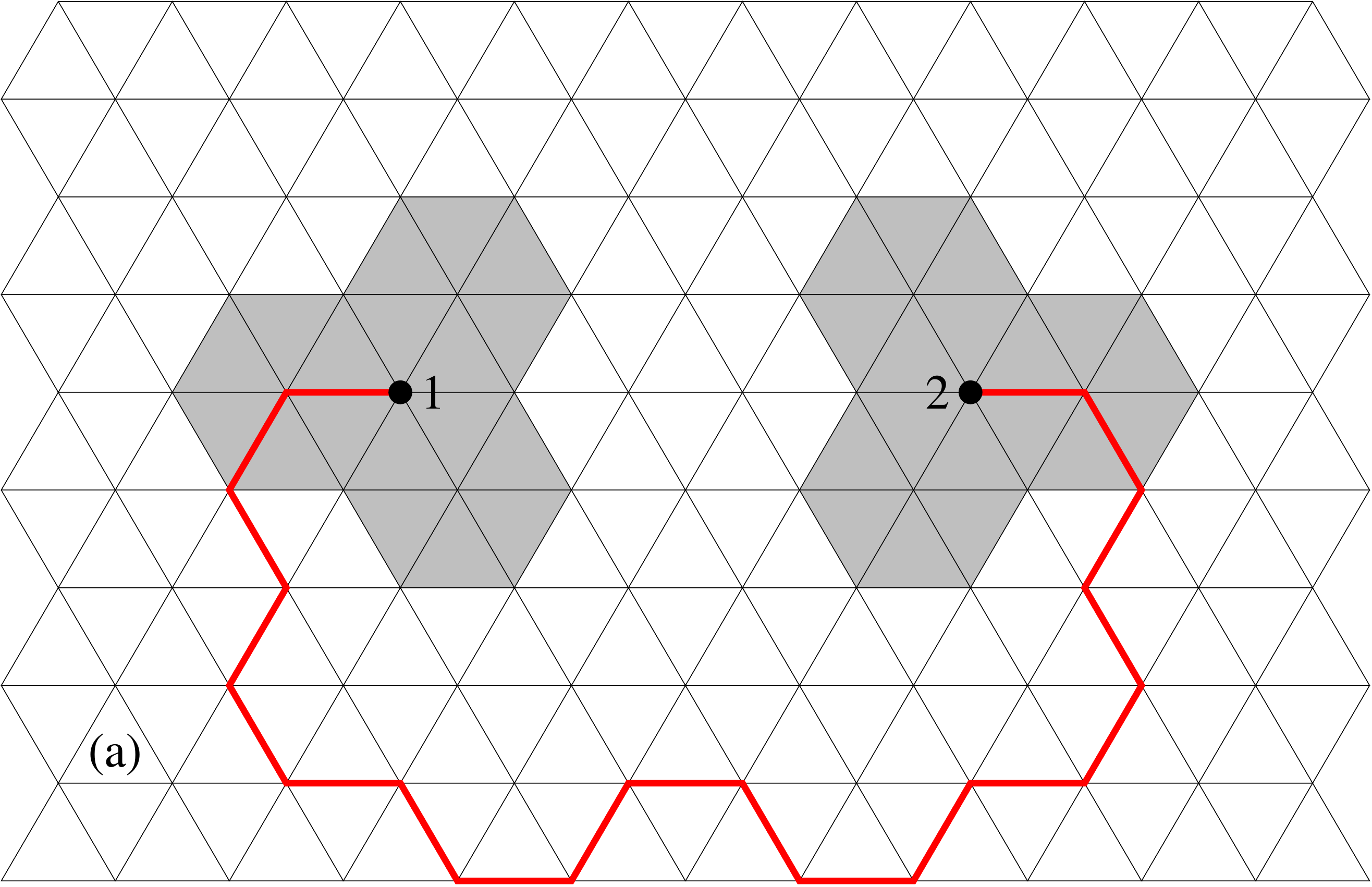}
\includegraphics[width=0.32\columnwidth]{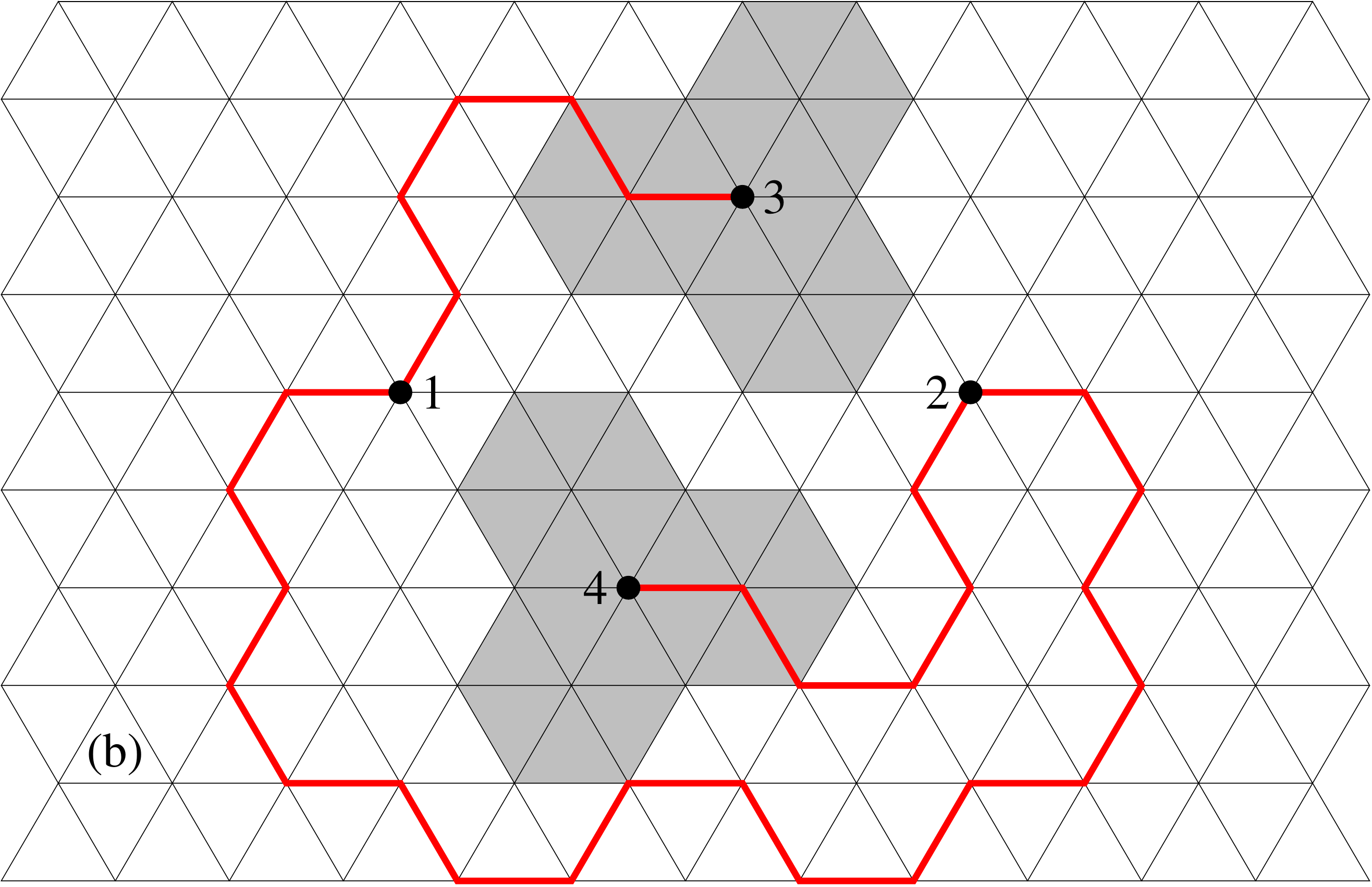}
\includegraphics[width=0.32\columnwidth]{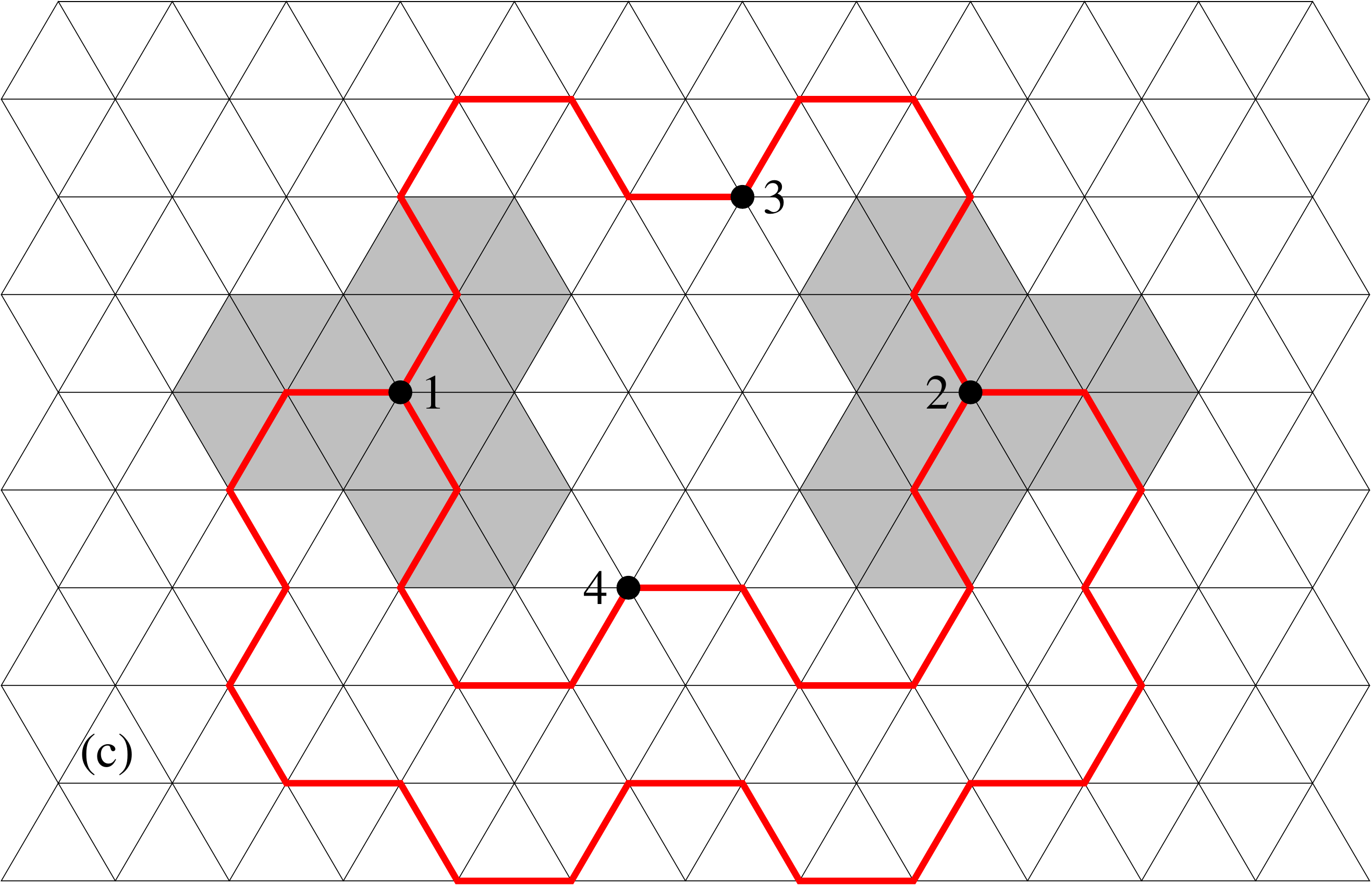}
\vskip 3mm
\includegraphics[width=0.32\columnwidth]{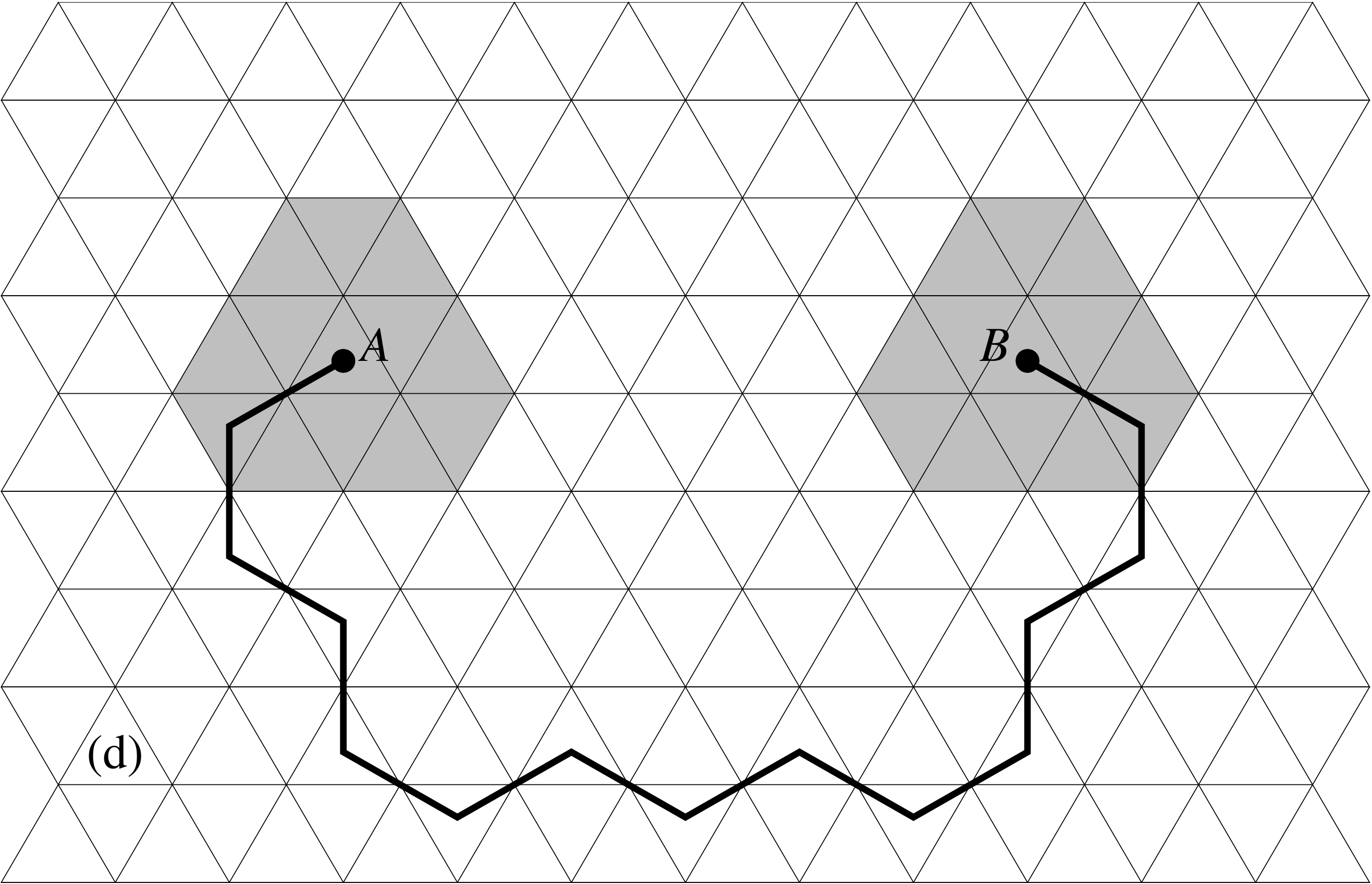}
\includegraphics[width=0.32\columnwidth]{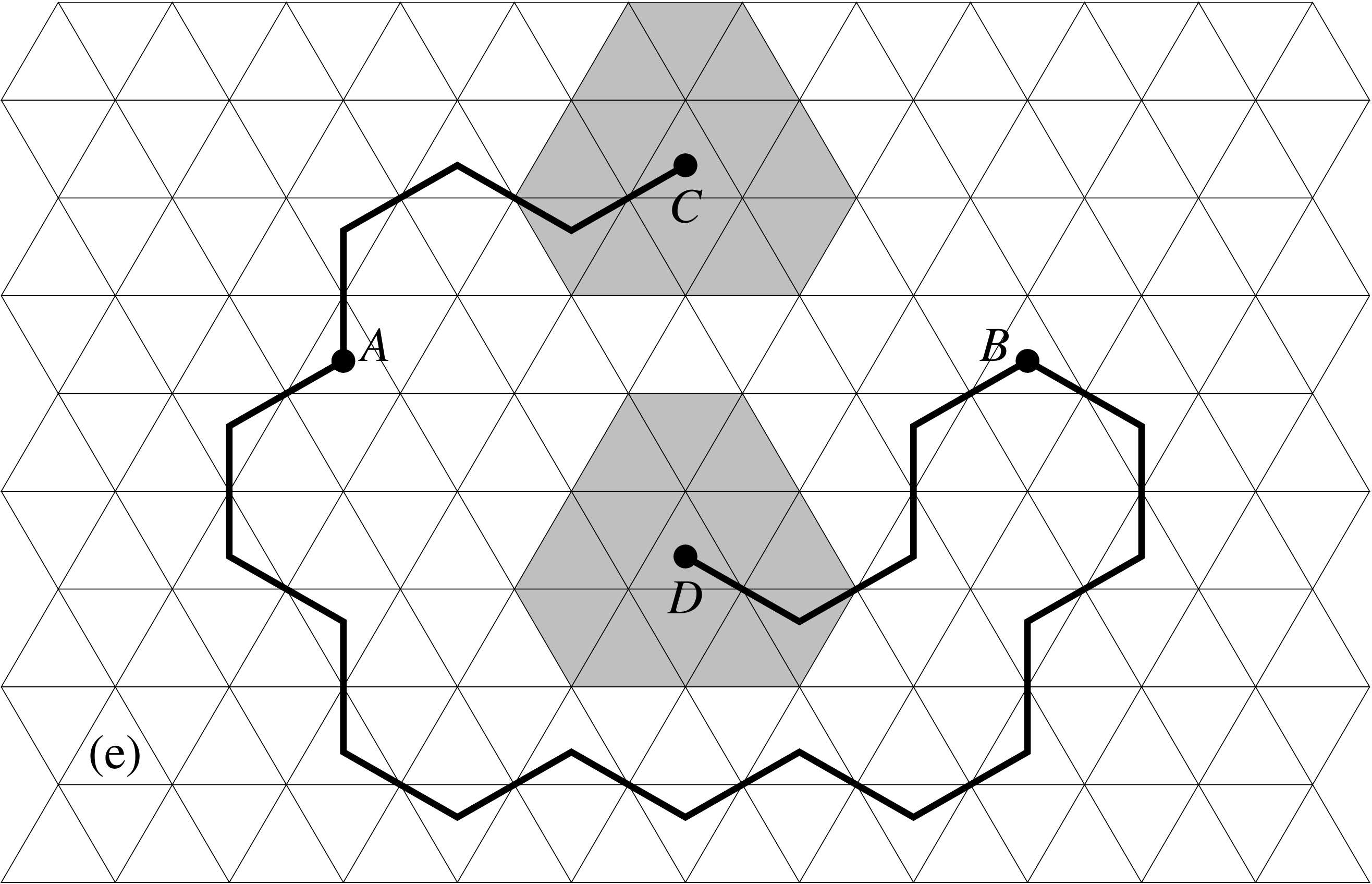}
\includegraphics[width=0.32\columnwidth]{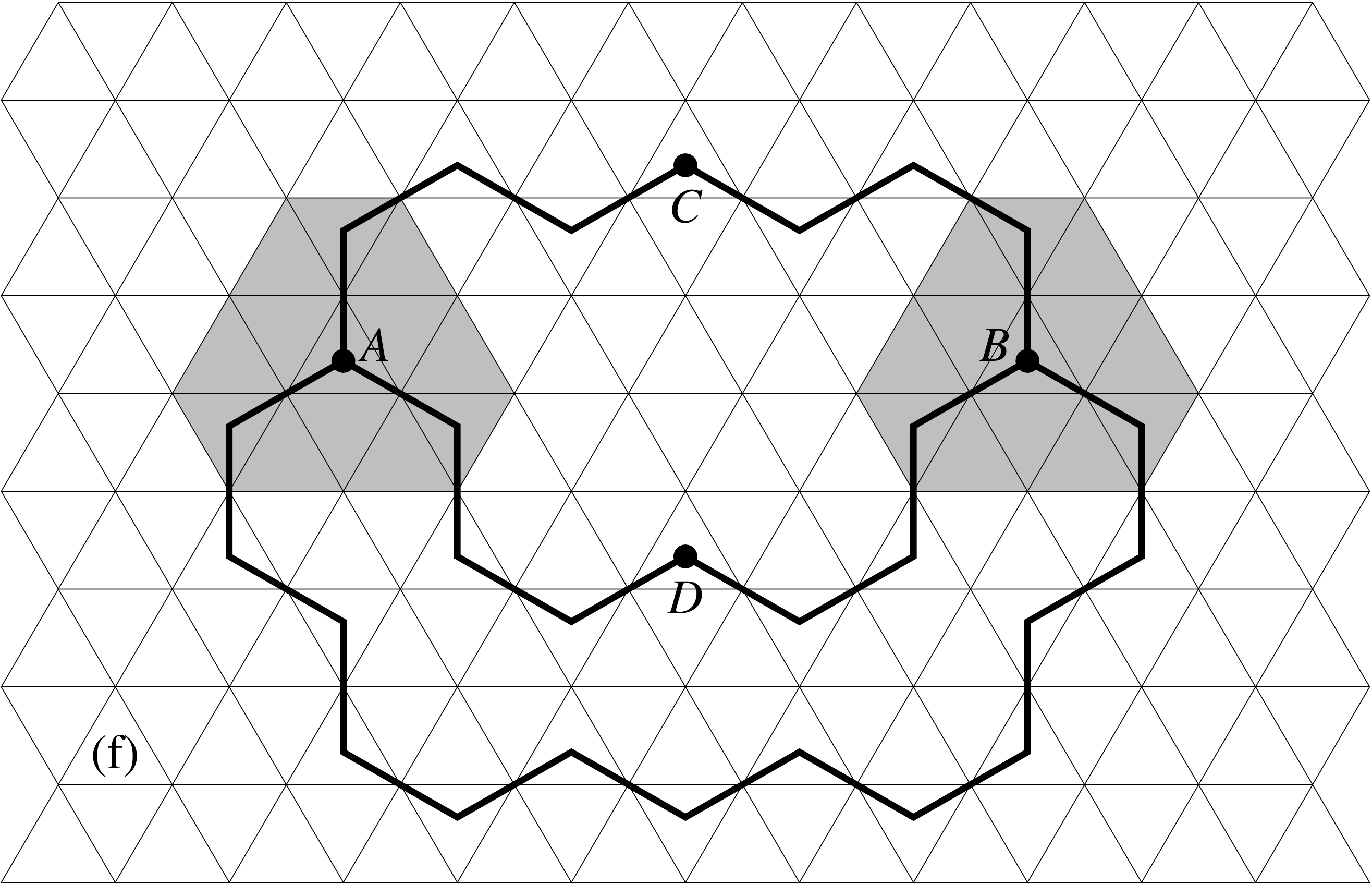}
\vskip 3mm
\includegraphics[width=0.32\columnwidth]{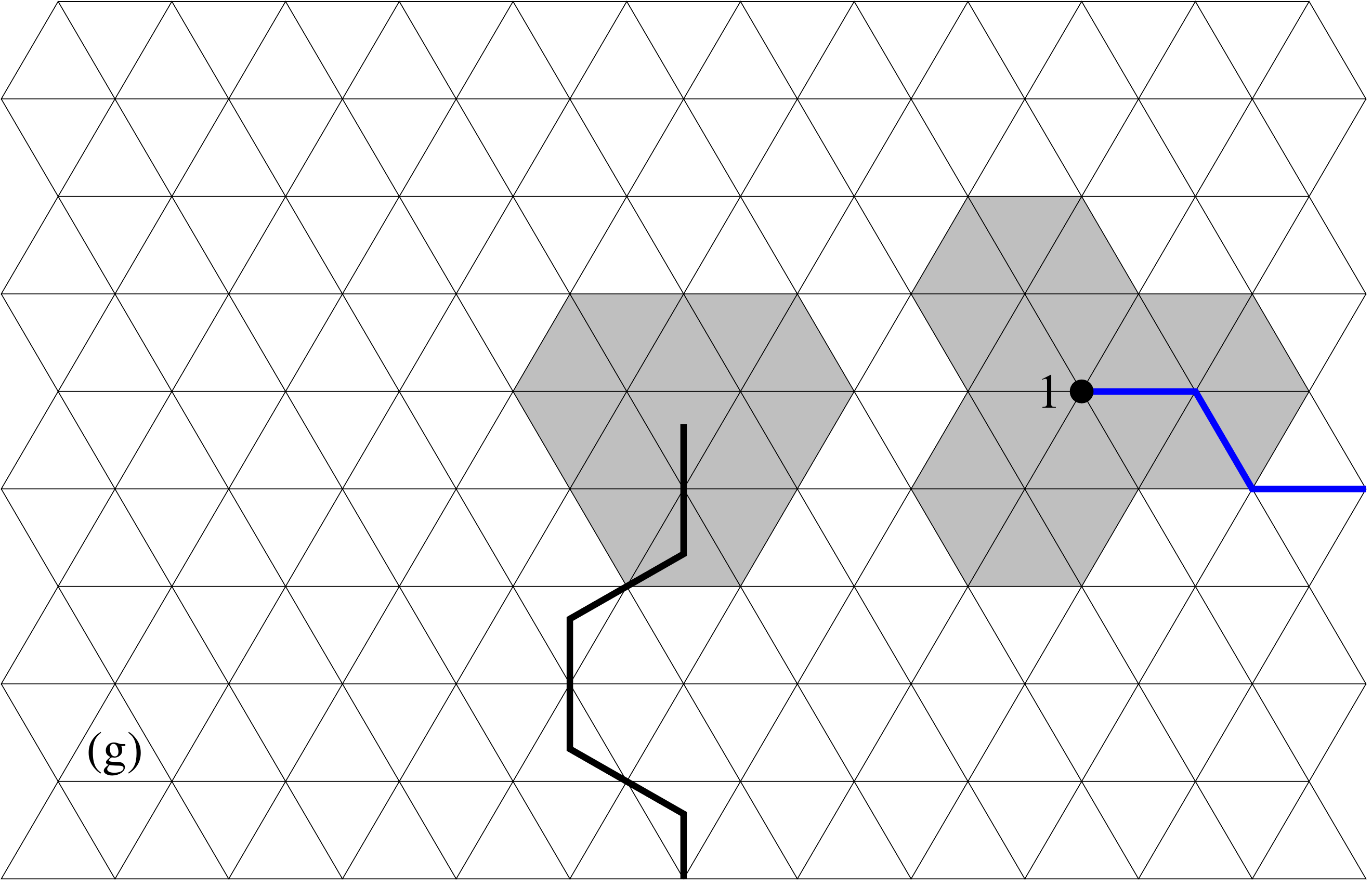}
\includegraphics[width=0.32\columnwidth]{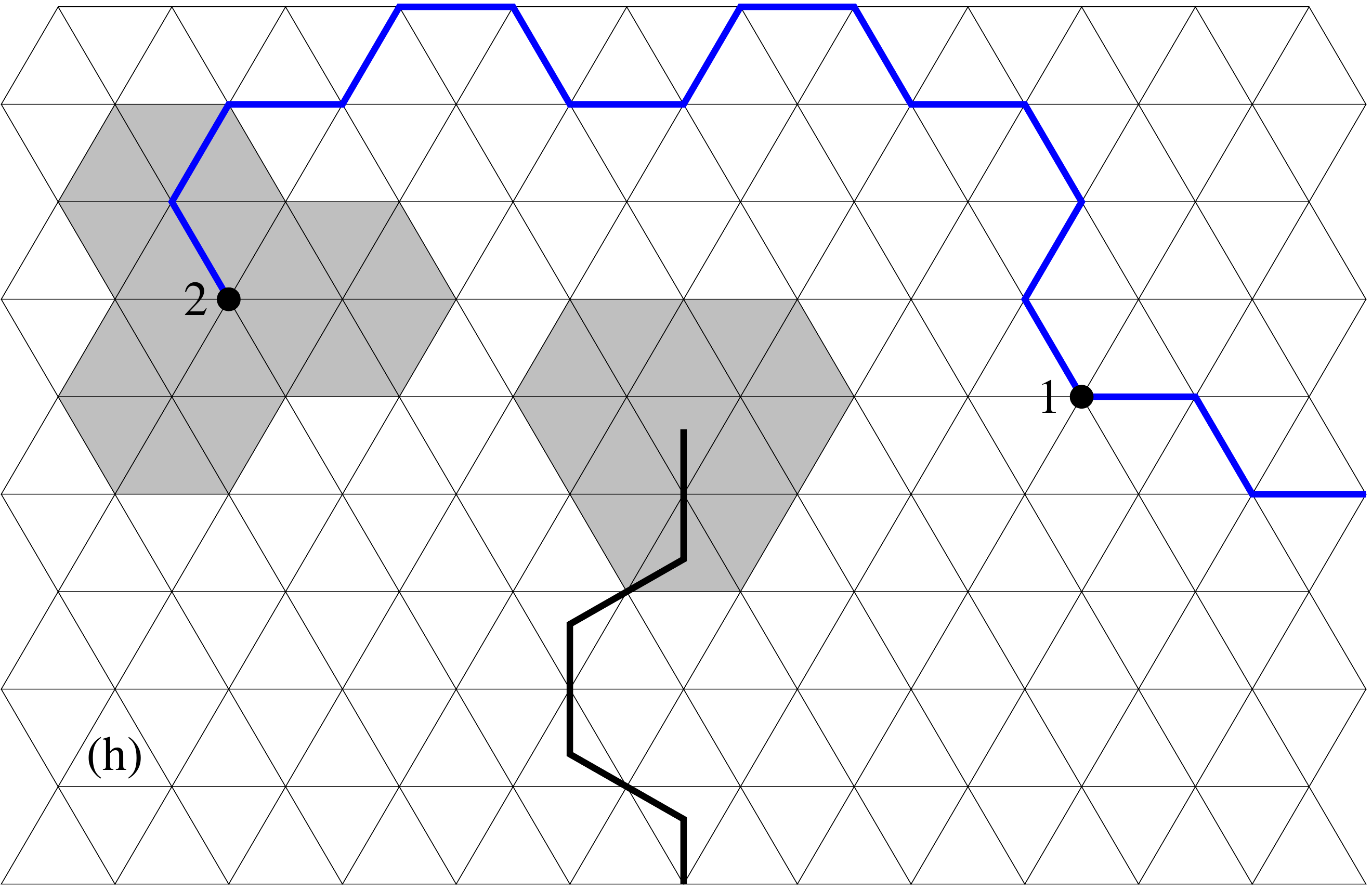}
\includegraphics[width=0.32\columnwidth]{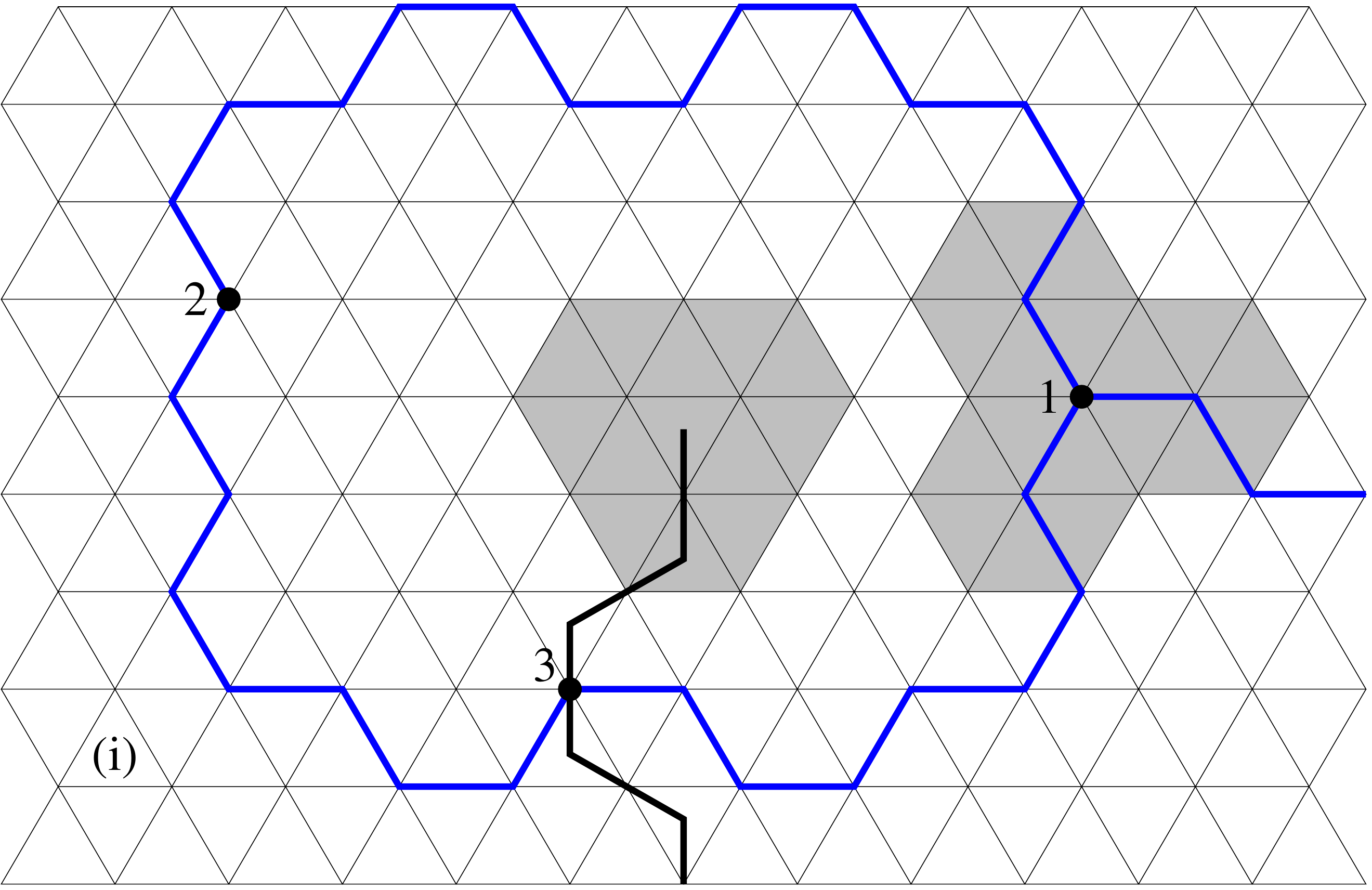}
\caption{(a)--(c) Braiding of two red fermions (the ends of a red string). The initial (a) and final (c) states are physically equivalent to each other. 1, 2, 3, and 4 are sites of the original lattice. (d)--(f) Braiding of two black bosons (the ends of a black string). The initial (d) and final (f) states are physically equivalent to each other. $A$, $B$, $C$, and $D$ are sites of a dual honeycomb. (g)--(i) A blue fermion (the end of a blue string) goes around a black boson (the end of a black string).}
\label{fig:exchange-FF-BB}
\end{figure}

\end{widetext}

\subsubsection{Fermions}

Quantum statistics of string ends can be established through braiding. Consider a state with two elementary particles obtained by the action of an open red string, Fig.~\ref{fig:exchange-FF-BB}(a),
\begin{equation}
|a\rangle = S_{21} |0\rangle.
\end{equation}
Here $21$ denotes a path running from 1 to 2. By extending the string on both ends, Fig.~\ref{fig:exchange-FF-BB}(b), we first obtain an intermediate state
\begin{equation}
|b\rangle = S_{42} S_{31} |a\rangle,
\end{equation}
and then eventually the state where the string ends have been braided clockwise, Fig.~\ref{fig:exchange-FF-BB}(c),
\begin{equation}
|c\rangle = S_{14} S_{23} S_{42} S_{31} |a\rangle,
\label{eq:c-from-a}
\end{equation}

States $|a\rangle$ and $|c\rangle$ are physically indistinguishable as they have the same set of excited hexagons near points 1 and 2. However, they may differ by a phase factor. Indeed, the operator connecting them in Eq.~(\ref{eq:c-from-a}) can be rearranged to form a Wilson loop,
\begin{equation}
|c\rangle = - S_{14} S_{42} S_{23}  S_{31} |a\rangle
	= - W_{14321} |a\rangle = - |a \rangle.
\end{equation}
The minus sign arises from the anti-commutation of strings $S_{42}$ and $S_{23}$ containing Pauli operators $\sigma_2^x$ and $\sigma_2^y$, respectively. The red Wilson loop $W_{14321} = +1$ in state $|a\rangle$ because there are no excited hexagons on the red honeycomb in that state.

We thus conclude that the ends of a string living on an original honeycomb (red, green, or blue) are fermions.

\subsubsection{Bosons}

In a similar way we braid the ends of a black string living on a dual honeycomb, Fig.~\ref{fig:exchange-FF-BB}(d)--(f). The initial state $|d\rangle$ and final state $|f\rangle$ are related by string extensions,
\begin{equation}
|f\rangle = S_{AD} S_{BC} S_{DB} S_{CA} |d\rangle.
\end{equation}
By rearranging the operators in the middle, we again reduce the product to a Wilson loop:
\begin{equation}
|f\rangle = S_{AD} S_{DB} S_{BC} S_{CA} |d\rangle
	= W_{ADBCA} |d\rangle = |d\rangle.
\end{equation}
This time, the operators $S_{DB}$ and $S_{BC}$ commute as they do not share common sites of the original lattice. The black Wilson loop $W_{ADBCA}$ in Fig.~\ref{fig:exchange-FF-BB}(f) is again trivial in the initial state $|d\rangle$ because it encloses no excited hexagons from its own honeycomb.

We thus find that ends of a string on a dual honeycomb (cyan, magenta, yellow, or black) are bosons.

\subsubsection{Mutual semions}

The ends of two strings of different types are of course distinguishable (e.g., with the aid of various Wilson loops), so exchanging them will result in a physically different state. Instead, we may move one of these particles around another, Fig.~\ref{fig:exchange-FF-BB}(g)--(i). The initial state $|g\rangle$ and the final state $|i\rangle$ are related by a blue Wilson loop,
\begin{equation}
|i\rangle = W_{1321} |g\rangle = - |g\rangle.
\end{equation}
Here the blue Wilson loop encloses one excited hexagon of the blue honeycomb created by the action of the black string terminating inside it. Put another way, the blue Wilson loop anti-commutes with the black open string because they intersect once at point 3, where they contribute Pauli operators $\sigma_3^y$ and $\sigma_3^z$, respectively.

Generally, whenever one particle encircles another particle of a different type, the quantum state acquires a factor of $-1$. Thus, different particles are mutual semions.

\subsection{Elementary excitations}
\label{sec:open-strings-elementary-excitations}

\begin{figure}
\includegraphics[width=0.95\columnwidth]{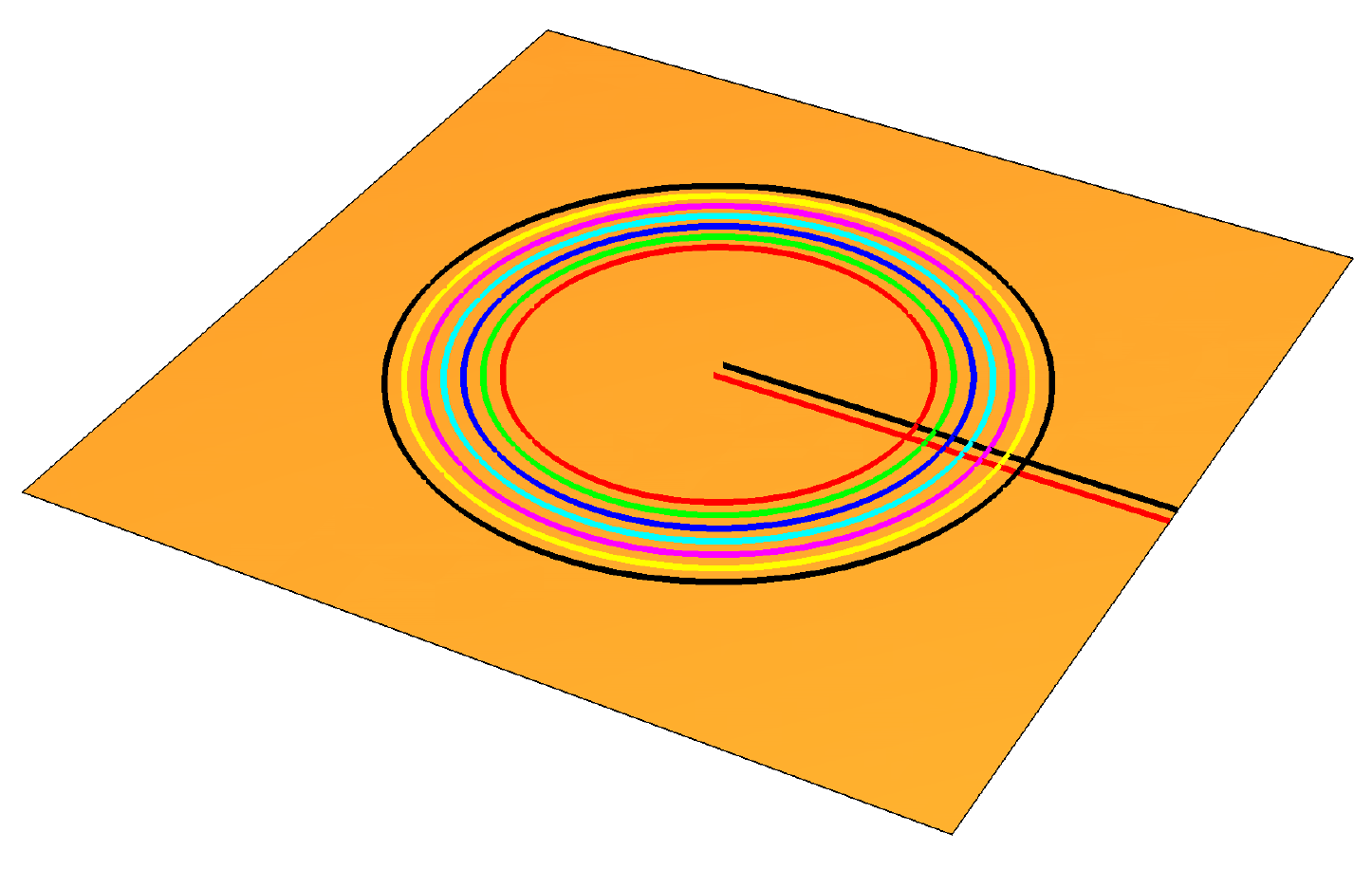}
\caption{Creation of a single elementary excitation $W = -1$ at the ends of two open strings detected by large Wilson loops.}
\label{fig:FB-strings-topological}
\end{figure}

In contrast to the $\z2$ spin liquids of Kitaev \cite{Kitaev2003} and Wen \cite{Wen2003}, the creation of an elementary excitation in our model requires the action of not one but at least two open strings. To see this from a topological perspective, consider a large lattice with no excitations, Fig.~\ref{fig:FB-strings-topological}. The absence of excitations can be confirmed by running 7 large Wilson loops (3 fermionic and 4 bosonic) that will have the values $+1$. Creating a single excited hexagon in the middle of these loops will change that. If the excited hexagon belongs to the red original honeycomb and to the black dual honeycomb then the red and black Wilson loops will switch to the value $-1$ and the rest will stay unchanged. To accomplish this, we can act on the ground state with two open strings, one red and one black, intersecting the Wilson loops once, Fig.~\ref{fig:FB-strings-topological}.

To construct an elementary excitation out of two strings on a microscopic level, we find it necessary to adjust slightly the definition of an open fermionic string (\ref{eq:open-F-string-def}). Instead of terminating a string on a site of the original lattice, we end it on a link:
\begin{equation}
\tilde{S}_\mathrm{T} =
	\sigma_n^\rho
	(\sigma_n^\nu \sigma_{n-1}^\nu)
	\ldots
	(\sigma_2^\beta \sigma_1^\beta)
	\sigma_1^\alpha.
\label{eq:open-F-string-def-alt}
\end{equation}
The two definitions of a fermionic string (\ref{eq:open-F-string-def}) and (\ref{eq:open-F-string-def-alt}) differ by local operators at the string ends, which preserves the braiding statistics of elementary particles.

\begin{figure}
\includegraphics[width=0.95\columnwidth]{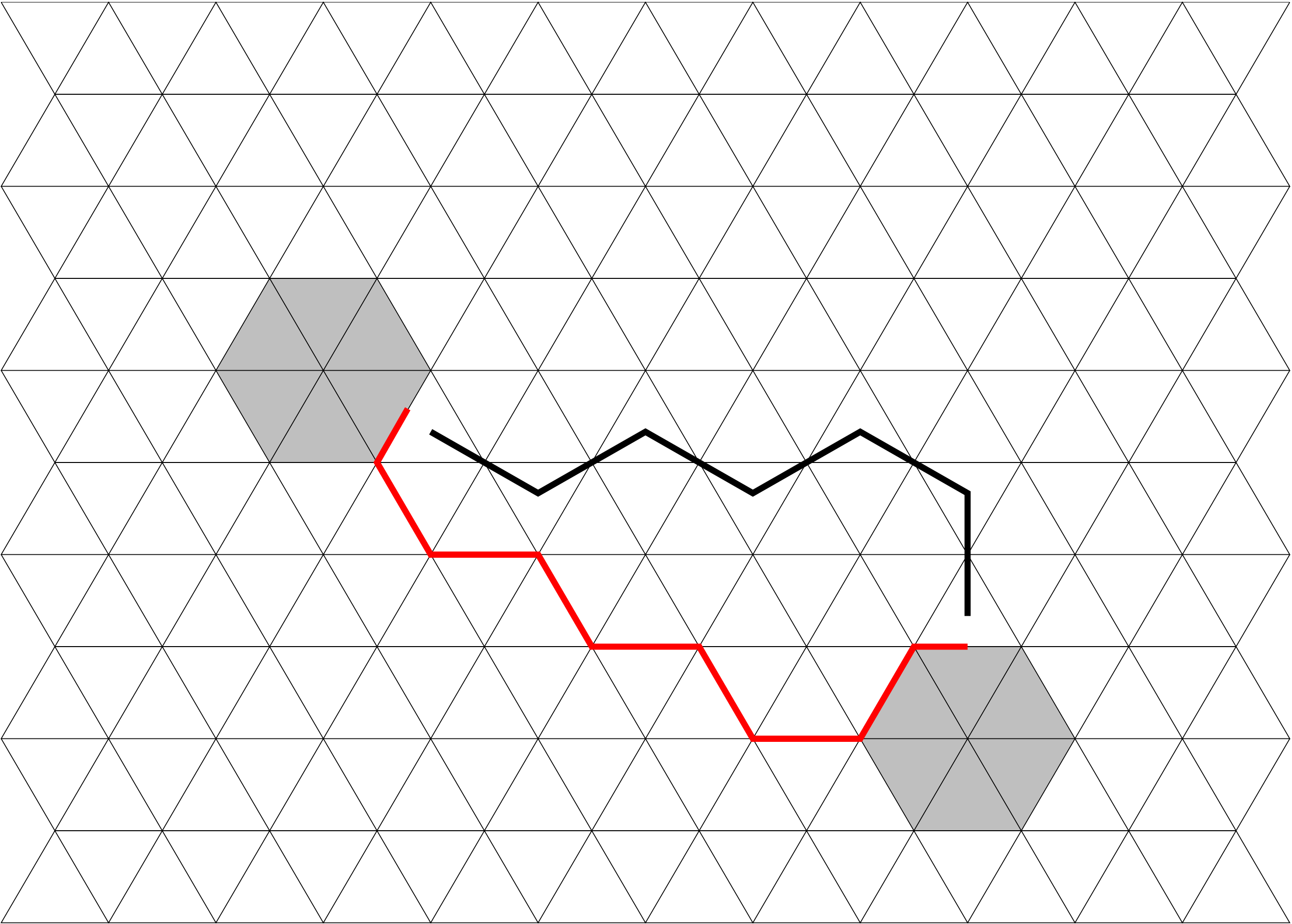}
\caption{The action of two open strings, bosonic $S_\tau$ (\ref{eq:open-B-string-def}) and fermionic $\tilde{S}_\mathrm{T}$ (\ref{eq:open-F-string-def-alt}), creates two elementary excitations $W = -1$ at the ends (shaded hexagons).}
\label{fig:FB-strings-micro}
\end{figure}

The combined action of a fermionic string $\tilde{S}_\mathrm{T}$ and a bosonic one $S_\tau$, produces a pair of elementary excitations $W = -1$, one at each end of the double string, Fig.~\ref{fig:FB-strings-micro}.

We thus find that elementary excitations $W = -1$ in our model come in $3 \times 4 = 12$ flavors determined by a choice of one fermionic ($R$, $G$, or $B$) and one bosonic ($C$, $M$, $Y$, or $K$) string. It is not possible to convert the an elementary excitation of one flavor (e.g., $RK$) into another (e.g., $BK$) without creating additional excitations.

\begin{figure}
\includegraphics[width=0.75\columnwidth]{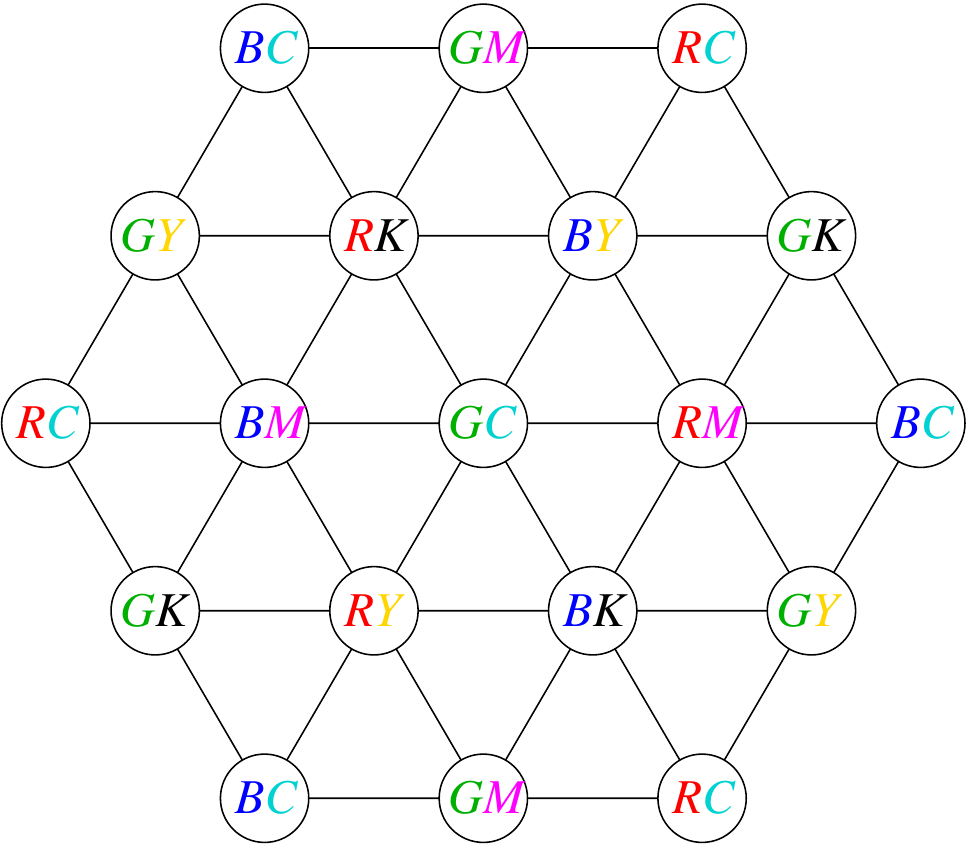}
\caption{Each hexagon, labeled by a circle at its center, is assigned two flavors, fermionic ($R$, $G$, or $B$) and bosonic ($C$, $M$, $Y$, or $K$). The resulting pattern of hexagons has a period $\sqrt{12} \times \sqrt{12}$.}
\label{fig:elementary-excitations}
\end{figure}

Note that excitations are created in pairs with the same flavors, e.g., $GC$ and $GC$. Hexagons with the same pairs of fermionic and bosonic flavors form a triangular lattice with the period $\sqrt{12} \times \sqrt{12}$, Fig.~\ref{fig:elementary-excitations}.

\section{Edge states}
\label{sec:edge}

\subsection{Majorana zero modes}
\label{sec:edge-Majorana}

As often happens with topological phases of matter, the edge of our quantum spin liquid harbors zero modes. In Wen's spin model \cite{Wen2003}, a straight edge of the square lattice has one Majorana zero mode per site. A similar count of the degrees of freedom in our model reveals two Majorana zero modes per site along a straight edge.

\begin{figure}
\includegraphics[width=0.95\columnwidth]{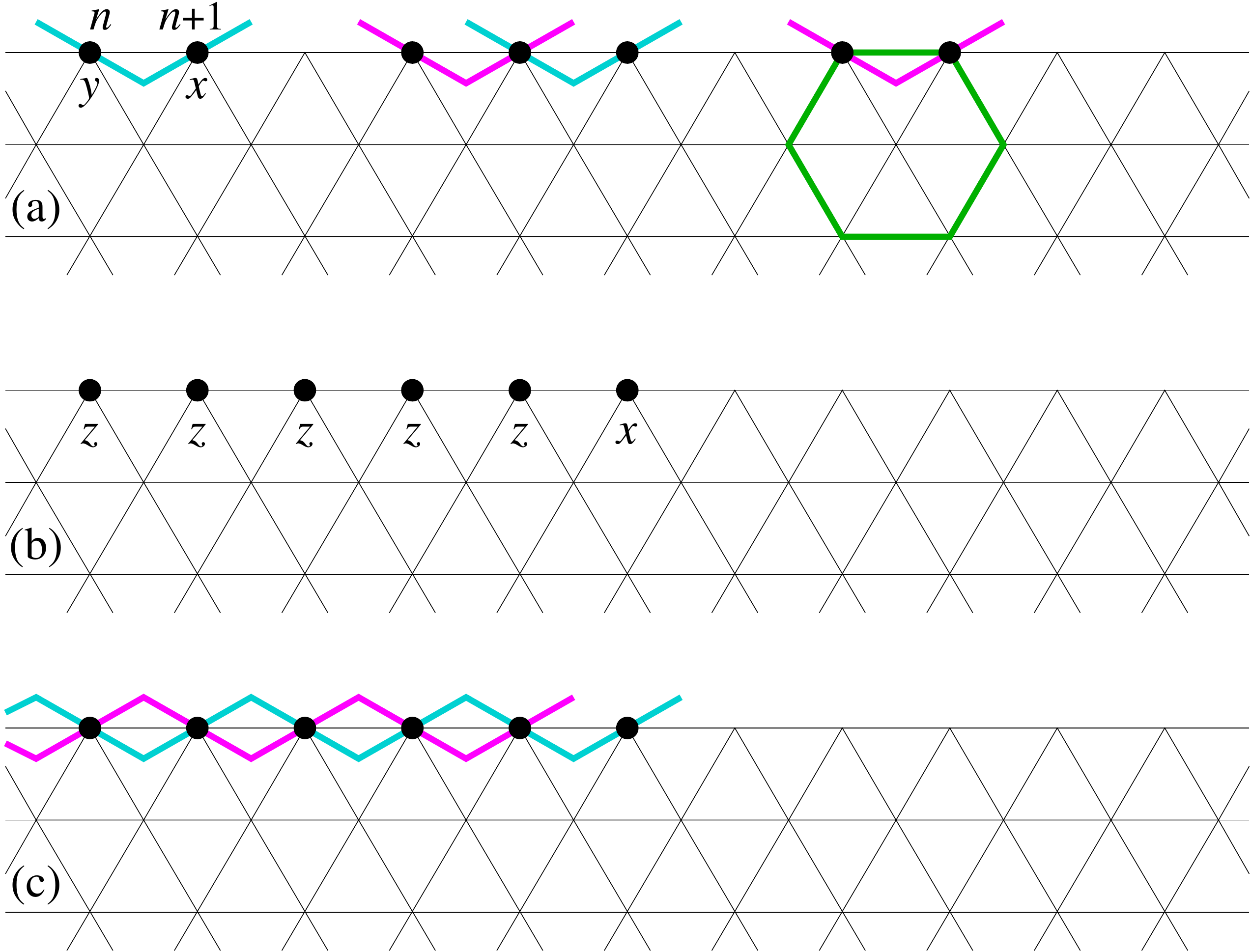}
\caption{(a) Short bosonic strings at the edge (\ref{eq:short-B-string}) commute with Wilson loops but may anticommute with one another. (b) A Majorana zero mode (\ref{eq:edge-Majorana}). (c) An equivalent combination of two bosonic strings (\ref{eq:open-B-string-def}).}
\label{fig:edge-modes}
\end{figure}

The existence of these zero modes manifests itself in an extensive degeneracy of the ground state (and generally of all energy levels) and can be established through the existence of multiple integrals of motion
\begin{equation}
S_{n,n+1} = \sigma_n^y \sigma_{n+1}^x,
\label{eq:short-B-string}
\end{equation}
in which one may recognize very short open bosonic strings (\ref{eq:open-B-string-def}) starting and ending just off the edge, Fig.~\ref{fig:edge-modes}(a). These short strings commute with the Hamiltonian (Eq.~\ref{eq:H-W}) but not necessarily with one other: strings of different flavors anticommute if they intersect once, Fig.~\ref{fig:edge-modes}(a). A large number of non-commuting integrals of motion indicates high degeneracy of energy levels.

The Majorana zero modes can be explicitly constructed as shown in Fig.~\ref{fig:edge-modes}(b). Operators
\begin{equation}
\alpha_{n+1} = \ldots \sigma_{n-2}^z \sigma_{n-1}^z \sigma_n^z \sigma_{n+1}^x
\label{eq:edge-Majorana}
\end{equation}
satisfy the standard algebra of Majorana fermions,
\begin{equation}
\alpha_{m} \alpha_n + \alpha_n \alpha_m = 2\delta_{mn}.
\end{equation}
One may worry that the string operator (\ref{eq:edge-Majorana}) represents a new string type overlooked in our earlier construction. (Its existence would increase the topological degeneracy.) Fortunately, that is not the case: the ``new'' string (\ref{eq:edge-Majorana}) is merely a product of two bosonic strings (\ref{eq:open-B-string-def}), of $C$ and $M$ flavors in Fig.~\ref{fig:edge-modes}(c). By combining two bosons that are mutual semions we obtain a fermion. The same happens in the toric code \cite{Kitaev2003}, where a combination of an electric charge and a flux (two bosons with mutual semion statistics) yields a fermion.

A second Majorana zero mode at the edge can be constructed from the other two bosonic strings.

\subsection{Majorana modes in a magnetic field}
\label{sec:edge-field}

\begin{figure}
\includegraphics[width=0.95\columnwidth]{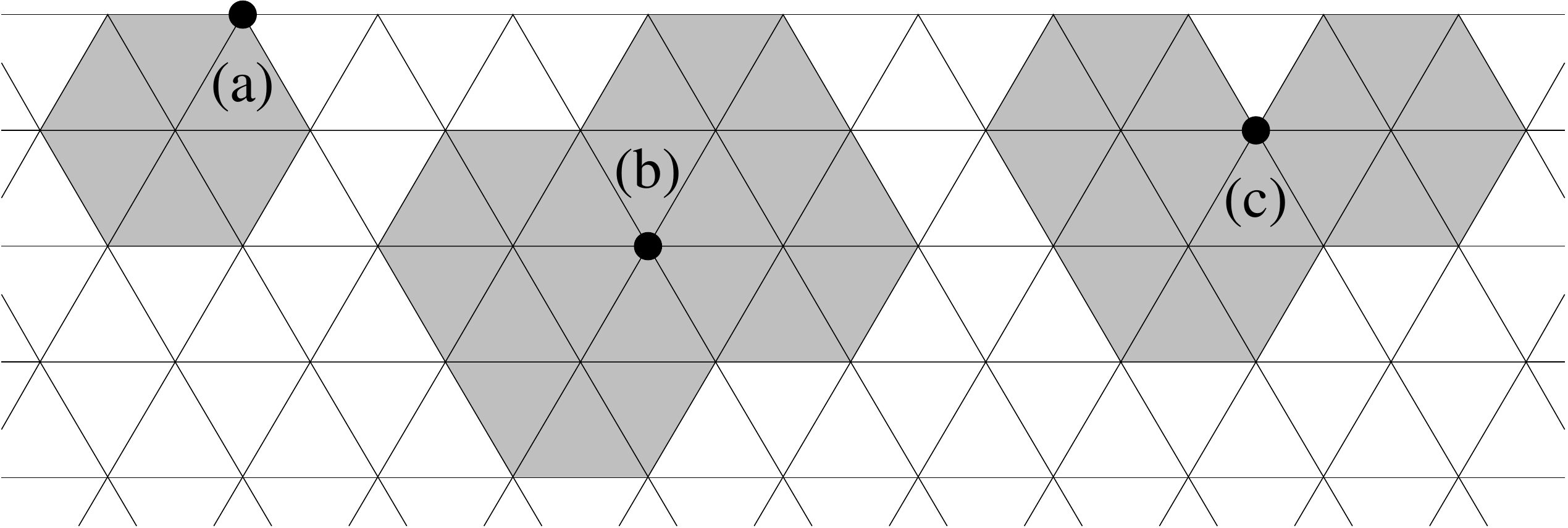}
\caption{The action of a $\sigma_n^x$ operator creates different numbers of excitations (shaded areas) depending on the location of site $n$ (black dot): (a) 1 hexagon for a site at a horizontal edge, (b) 4 hexagons for a site in the bulk (excluding the hexagon centered on site $n$), (c) 3 hexagons for a site near a horizontal edge.}
\label{fig:edge-modes-field}
\end{figure}

Zero modes at the edge of a spin liquid are sensitive to perturbations \cite{Kitaev2006}. The Majorana modes acquire a dispersion and propagate along the edge. We have analyzed their response to a weak uniform magnetic field,
\begin{equation}
H_1 = - \sum_n (h_x \sigma_n^x + h_y \sigma_n^y + h_z \sigma_n^z),
\end{equation}
along the lines of \textcite{Kou2013}.

The application of a perturbation term such as $- h_x \sigma_n^x$ creates 4 excited hexagons in the bulk, Fig.~\ref{fig:edge-modes-field}. To the second order in $\mathbf h$, the perturbation simply shifts the energy of the ground state by
\begin{equation}
\Delta E_{\mathrm{bulk}}^{(2)} = - \frac{h_x^2 + h_y^2 + h_z^2}{8}
\end{equation}
for every bulk site. For a site near a horizontal edge, perturbations $- h_x \sigma_n^x$ and $- h_y \sigma_n^y$ creates 3 excitations (Fig.~\ref{fig:edge-modes-field}), whereas $- h_z \sigma_n^z$ just 2; the energy of the ground state is shifted by
\begin{equation}
\Delta E_{\mathrm{near\ edge}}^{(2)} = - \frac{h_x^2 + h_y^2}{6} - \frac{h_z^2}{4}
\end{equation}
per site. For a site at the edge, the energy shift is
\begin{equation}
\Delta E_{\mathrm{edge}}^{(2)} = - \frac{h_x^2 + h_y^2}{2} - \frac{h_z^2}{4}.
\end{equation}

In addition to the trivial shift of the energy, the second-order perturbation generates the following virtual processes: an excited hexagon can be at first created by $- h_x \sigma_{n+1}^x$ and then destroyed by $- h_y \sigma_n^y$, or vice versa. These virtual processes give rise to an effective perturbation Hamiltonian for a straight edge, Fig.~\ref{fig:edge-modes}(a):
\begin{equation}
H^{(2)}_\mathrm{eff} = -h_x h_y \sum_n \sigma_n^y \sigma_{n+1}^x
\end{equation}
made of short bosonic strings (\ref{eq:short-B-string}) and thus commuting with the bulk Hamiltonian (\ref{eq:H-W}).

Expressed in terms of the Majorana fermions (\ref{eq:edge-Majorana}), the effective Hamiltonian reads
\begin{equation}
H^{(2)}_\mathrm{eff} = -h_x h_y \sum_n i \alpha_n \alpha_{n+1}.
\end{equation}
The excitation spectrum of the edge fermions is
\begin{equation}
\epsilon_k = 4 |h_x h_y| \sin{k},
\quad
0 \leq k \leq \pi.
\end{equation}

For an open cylinder with $N$ sites and two straight edges oriented as in Fig.~\ref{fig:edge-modes-field} and containing $L$ sites each, the second-order correction to the energy of the ground state can be split into the bulk and edge parts,
\begin{equation}
\Delta E^{(2)} = - N\frac{h_x^2 + h_y^2 + h_z^2}{8} + 2 E_\mathrm{edge},
\label{eq:E2}
\end{equation}
where the edge energy includes the trivial local shift and the zero-point energy of Majorana fermions propagating along the edge:
\begin{equation}
E_\mathrm{edge} = - \frac{5 h_x^2 + 5 h_y^2 + 3 h_z^2}{12} L
	-  2 |h_x h_y| \sum_{0<k<\pi} \sin{k}.
\label{eq:E2-edge}
\end{equation}

Periodic boundary conditions for edge spins, $\sigma_L^\alpha = \sigma_0^\alpha$, translate into either periodic or antiperiodic boundary conditions for Majorana fermions,
\begin{equation}
\alpha_L = - \alpha_0 W,
\end{equation}
where
\begin{equation}
W = \sigma_1^z \sigma_2^z \ldots \sigma_L^z = W_C W_M
\end{equation}
is a global Wilson loop winding around the cylinder.

For $W = +1$, the boundary conditions are antiperiodic and Majorana fermions have momenta $k_n = 2\pi (n-1/2)/L$. For $W=-1$, $k_n = 2\pi n/L$. In both cases, $n = 1, 1, \ldots, L$. We assume that $L$ is even. The edge energy in the two sectors is
\begin{eqnarray}
E_\mathrm{edge}^{W=+1} = - \frac{5 h_x^2 + 5 h_y^2 + 3 h_z^2}{12} L
		- 2 |h_x h_y| \csc{\frac{\pi}{L}},
\\
E_\mathrm{edge}^{W=-1} = - \frac{5 h_x^2 + 5 h_y^2 + 3 h_z^2}{12} L
		- 2 |h_x h_y| \cot{\frac{\pi}{L}}.
\end{eqnarray}
The state with a trivial global flux $W=+1$ is lower in energy:
\begin{equation}
E_\mathrm{edge}^{W=+1} - E_\mathrm{edge}^{W=-1}
	= - 2 |h_x h_y| \tan{\frac{\pi}{2L}} < 0.
\end{equation}

\section{Numerical diagonalization}
\label{sec:numerics}

\begin{figure}
\includegraphics[width=0.95\columnwidth]{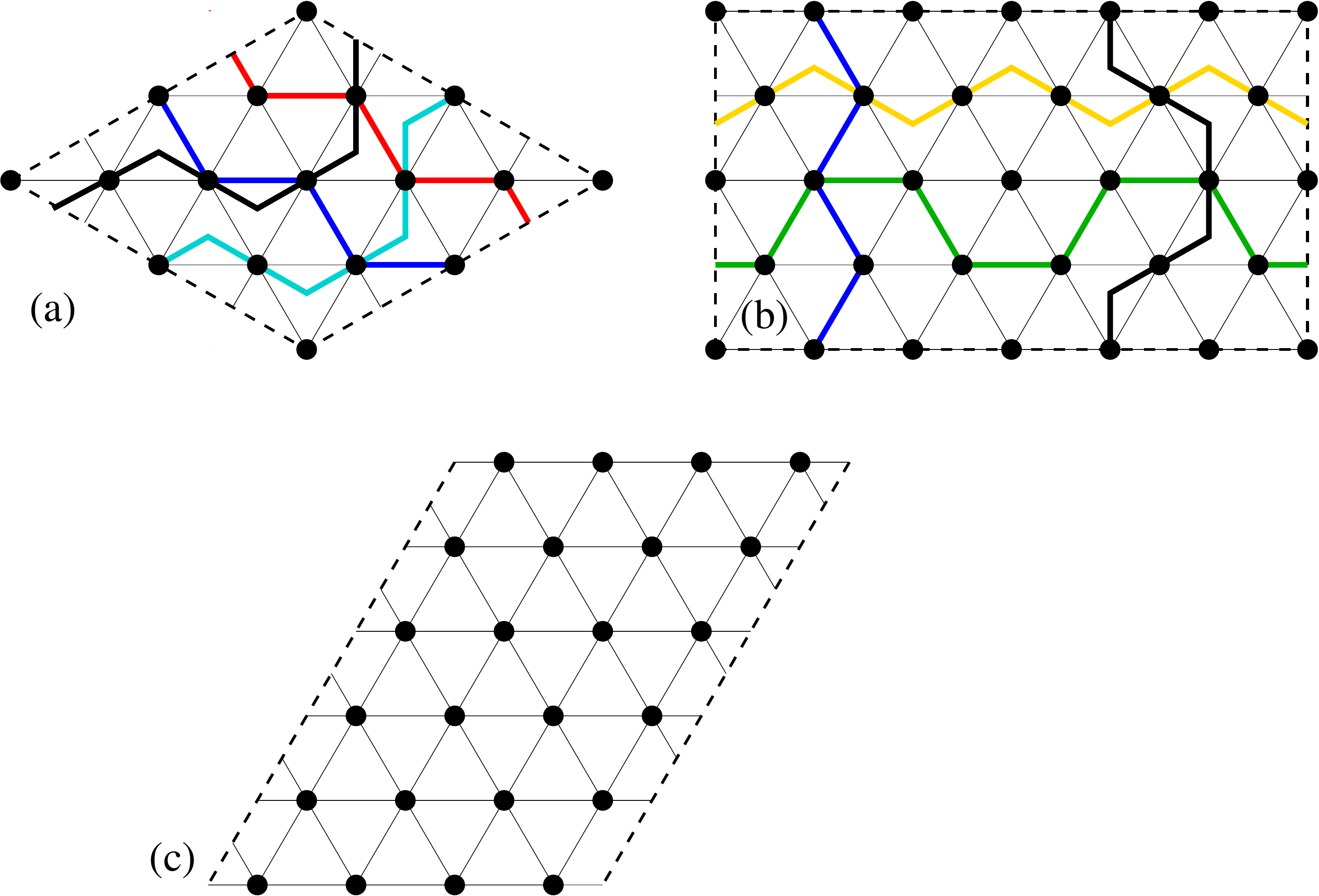}
\caption{(a) and (b) Clusters with periodic boundary conditions (opposite edges identified) containing $N = 12$ (a) and 24 (b) sites. Some of the global Wilson loops are shown. (c) A cluster with $N = 24$ sites and two open edges (top and bottom) containing $L = 4$ sites each. Left and right edges are identified.}
\label{fig:numerical-tori}
\end{figure}

To verify our results, we obtained energy spectra for some finite clusters. We employed brute-force diagonalization of the Hamiltonian (\ref{eq:H-W}) for small clusters; for larger ones, we used the Lanczos algorithm and took into account translational symmetry to reduce the sizes of the Hilbert spaces. Anticipating high topological degeneracy, we seeded the Lanczos algorithm multiple times and orthogonalized low-energy states with respect to the ones already obtained.

\subsection{Topological degeneracy and elementary excitations}

To check the predicted topological degeneracy of 64 on a torus, we examined the energy spectra of two clusters containing $N = 12$ and 24 sites with periodic boundary conditions compatible with the existence of 7 distinct types of strings, Fig.~\ref{fig:numerical-tori}. Both clusters have 64-fold degenerate ground states with energies $E_0 = -N$, as expected. Here we label energy levels by the number of excited hexagons.

The first excited state of the $N = 24$ cluster has the energy $E_2 = -20 = E_0 + 4$, which corresponds to 2 excited hexagons. This energy level had the degeneracy of  $768 = 12 \times 64$, of which the factor of 64 is the topological degeneracy. The remaining factor of 12 reflects the number of distinct excitation types in this cluster. As explained in Sec.~\ref{sec:open-strings-elementary-excitations}, excitations are created in pairs at hexagons belonging to one of the 12 sublattices. In the cluster with $N = 24$ sites, there is exactly one pair of hexagons for each of the 12 sublattices.

The $N=12$ cluster has only one hexagon for each pair of flavors. Thus it is not possible to create a pair of excitations out of the ground state. Indeed, this cluster's energy spectrum does not have levels with energy $E_2 = E_0 + 4$. The lowest excited state with energy $E_4 = E_0 + 8$ contains 4 excited hexagons in combinations of two fermionic and two bosonic flavors, e.g., $RY$, $RC$, $BY$, $BC$. The number of possible combinations is $\frac{3!}{2! \cdot 1!} \frac{4!}{2! \cdot 2!} = 18$, so we expect the total degeneracy of $18 \times 64 = 1152$, which is indeed the case.

\subsection{Edge modes}

The energy of the ground state of a cluster with $N = 24$ sites and two open edges with $L = 4$ sites each, shown in Fig.~\ref{fig:numerical-tori}(c), has the following dependence on the applied magnetic field $\mathbf h = (\frac{h}{\sqrt{3}},\frac{h}{\sqrt{3}},\frac{h}{\sqrt{3}})$ of strength $h < 0.02$:
\begin{equation}
E_0(h) = E_0(0) - 7.77452 h^2.
\end{equation}
This numerical result matches well the second-order correction for the $W=+1$ state (Sec.~\ref{sec:edge-field}):
\begin{equation}
\Delta E^{(2)} = -\left( \frac{53}{9} + \frac{4\sqrt{2}}{3}\right) h^2 \approx - 7.77451 h^2.
\end{equation}
The agreement confirms the existence of propagating Majorana modes at the edges in the presence of an external magnetic field.

\section{Discussion}
\label{sec:discussion}

We have presented an exactly solvable model of a quantum spin liquid on a triangular lattice with six-spin interactions. Strong quantum fluctuations generate long-range entanglement of spins and topological order. Elementary excitations are nonlocal objects. To understand their nature, we have constructed natural building blocks of the model---string operators defined on links of either the original or dual lattice. The geometry of our model gives rise to a larger variety of strings than in predecessor square-lattice models of \textcite{Kitaev2003} and \textcite{Wen2003}. Both of those had 2 bosonic strings and 1 fermionic, whereas ours has 4 bosonic and 3 fermionic string types. In all of these models, ends of strings are associated with elementary particles (hence the designation of strings as bosonic or fermionic). Particles of two distinct types are mutual semions, a feature also found in the Kitaev and Wen models.

Elementary excitations in our model (defined as smallest quanta of energy) are distinct from elementary particles (ends of strings). A single excitation can be viewed as a pair of elementary particles, one boson and one fermion. We thus have a large number, $4 \times 3 = 12$, distinct types of elementary excitations living on 12 sublattices, Fig.~\ref{fig:elementary-excitations}. These elementary excitations are static in the exactly solvable model. A small modification of the Hamiltonian (\ref{eq:H-W}) will make them mobile. For weak perturbations away from the solvable point, these excitations will only be able to tunnel between sites of their own sublattice separated distance $\sqrt{12}$ apart.

The large number of string types (and of elementary particles) directly translates into high topological degeneracy, $2^6$ on a torus. This number, and the Abelian nature of the anyons, suggests that our model is equivalent to three decoupled $\z2$ gauge fields, each of which contributes a factor of $2^2$. This could be verified by constructing three pairs of $\z2$ electric charges and fluxes $\{e_i,m_i\}$, $i=1,2,3$, where the bosonic electric charge $e_i$ and magnetic flux $m_i$ within any pair would be mutual semions and would have trivial braiding statistics with the members of the other pairs. Although this construction is indeed possible, it requires the use of composite particles as there are only 4 elementary bosons. E.g., $\{e_1, m_1\} = \{C,M\}$, $\{e_2,m_2\} = \{RY,RK\}$, $\{e_3,m_3\} = \{BCM,GCM\}$. Such an asymmetric construction does not look natural and provides no additional insights. If anything, it obscures the link between lattice symmetries and topological order \cite{Kitaev2006, PhysRevB.87.104406}.

It would be interesting to see whether one may find a model with similar properties but less contorted interactions than the six-spin term (\ref{eq:W}). That this is possible, at least in principle, can be seen from the example of the toric code \cite{Kitaev2003, Wen2003}. Its universal features---the topological order and anyon excitations---are reproduced in the gapped phase of Kitaev's honeycomb model with more realistic two-spin interactions \cite{Kitaev2006}.

\section*{Acknowledgments}

We thank the Johns Hopkins Homewood High-Performance Computing Cluster for making available its computational resources. This research was supported in part by the U.S. Department of Energy, Office of Basic Energy Sciences, Division of Materials Sciences and Engineering under Award DE-FG02-08ER46544, and by Perimeter Institute for Theoretical Physics. Research at Perimeter Institute is supported by the Government of Canada through the Department of Innovation, Science and Economic Development Canada and by the Province of Ontario through the Ministry of Research, Innovation and Science.

\bibliographystyle{apsrev4-1}
\bibliography{project6}

\end{document}